\documentclass{JHEP3}

\usepackage{epsfig}
\usepackage{amsfonts,amssymb,amsmath,amsthm}

\newcommand{\be}{\begin{equation}}
\newcommand{\ee}{\end{equation}}

\newcommand{\ba}{\begin{eqnarray}}
\newcommand{\ea}{\end{eqnarray}}

\newcommand{\half}{\ensuremath{\frac{1}{2}}}

\newtheorem*{claim}{Claim}
\newtheorem*{theorem}{Theorem}
\newcommand{\tr}{\operatorname{Tr}}
\newcommand{\tn}{\tilde{n}}

\title{Five-Brane Thermodynamics from the Matrix Model}

\author{
Kazuyuki Furuuchi$^{\dagger \ddagger}$, 
Ehud Schreiber$^{\dagger \ddagger}$ and
Gordon W. Semenoff$^\dagger$\\
$^\dagger$ Department of Physics and Astronomy,
University of British Columbia,\\ 
6224 Agricultural Road, Vancouver, BC, V6T 1Z1, Canada.\\
$^\ddagger$ Pacific Institute for the Mathematical Sciences, 
University of British Columbia,\\ 
1933 West Mall, Vancouver, BC, V6T 1Z2, Canada.
\email{E-mail:furuuchi,schreiber,semenoff$@$physics.ubc.ca}
\thanks{Work supported in part by NSERC of Canada and the PIMS String Theory 
CRG.}
}

\abstract{A certain sector of the matrix model for M-theory on a plane wave 
background has recently been shown to produce the transverse five-brane.  
We consider this theory at finite temperature. We find that, at a critical 
temperature it has a Gross-Witten phase transition which corresponds to 
deconfinement of the matrix model gauge theory. We interpret the phase 
transition as the Hagedorn transition of M-theory and of type II string theory
in the five-brane background. We also show that there is no Hagedorn behaviour
in the transverse membrane background case.}

\preprint{}


\begin{document}
\section{Introduction and Summary}

The matrix theory conjecture states that the large N limit of a certain matrix quantum mechanics is an exact description of light-cone quantized M-theory in flat eleven-dimensional space-time \cite{Banks:1996vh}
(for a review see \cite{Taylor:2001vb}). Many of the objects of M-theory, such as the supergraviton, supermembrane and M five-brane emerge in a natural way from the matrix model and some of their low-energy interactions have been computed and matched with supergravity.  However, the study of this seemingly simple model
has been plagued by its nonlinearity and the existence of flat directions in the potential leading to a continuous spectrum \cite{deWit:1988ig}.

Recently a matrix model of M-theory in an 11-dimensional plane-wave geometry has been studied~\cite{Berenstein:2002jq}-\cite{Maldacena:2002rb}.   It has the great advantage that one can formulate an accurate perturbative expansion of the model.  Properties of states and symmetries have been analyzed in this limit \cite{Dasgupta:2002hx}-\cite{Kim:2002if}.  It has also been realized that some of the states are protected by supersymmetry.

The plane-wave background has a single dimensional parameter $\mu$ which appears
in the metric and the constant four-form field strength,
\begin{eqnarray}
\label{pmetric}
ds^2&=&-2dx^+dx^- - \left( \left(\frac{\mu}{3}\right)^2(x^a)^2+
\left(\frac{\mu}{6}\right)^2(x^{a'})^2\right)(dx^+)^2+(dx^a)^2+(dx^{a'})^2
\nonumber \\
F_{+123}&=&\mu
\end{eqnarray}
This is a 1-parameter family of backgrounds which goes to Minkowski space when $\mu\to 0$.
The  matrix model for M-theory on this background has
action \cite{Berenstein:2002jq}
\begin{eqnarray}
S=R\int dt\tr\left\{ \frac{1}{2R^2}\left( D_t X^i\right)^2
+\frac{i}{R}\psi^{t}D_t\psi+\psi^t\Gamma^i\left[ X^i,\psi\right]+
\frac{1}{4}\left[ X^i,X^j\right]^2 \right.\nonumber \\
\left.
-\frac{1}{2}\left(\frac{\mu}{3R}\right)^2(X^a)^2-\frac{1}{2}\left( \frac{\mu}{6R}\right)^2(X^{a'})^2
-i\frac{\mu}{4R}\psi^t\Gamma^{123}\psi-i\frac{\mu}{3R}\epsilon_{abc}X^aX^bX^c\right\}
\label{mmaction}
\end{eqnarray}
We have set the M-theory Planck length to one, $\ell_p=1$, and the conventions
for indices are $i,j,...=1,\ldots,9~$ and $~~a,b,\ldots=1,2,3$,
$~~a',b',\ldots=4,\ldots,9$.
All degrees of freedom are $N\times N$ Hermitian matrices. The covariant
derivative is
\begin{equation}
D_t X=\frac{d}{dt}X+i\left[ A,X \right]
\end{equation}
This matrix model is conjectured to be a discrete light-cone quantization (DLCQ) of M-theory
where $R$ is the compactification radius of the null coordinate $x^-$ and
there are $N$ units of light-cone momentum, $p^+=N/R$.  It is a
1-parameter deformation of the BFSS matrix model \cite{Banks:1996vh}
which described M-theory on Minkowski space.  Just as the BFSS model
could also be regarded as the effective action describing the low energy
dynamics of a collection of $N$ D0-branes of type IIA string theory on
Minkowski space, with the appropriate re-interpretation of the
parameters, (\ref{mmaction}) could also describe D0-branes in a
background 
of type IIA string theory.

The matrix model can be analyzed in perturbation theory.   As was noted
in ref.\ \cite{Berenstein:2002jq}, one interesting feature of the semiclassical expansion is the existence of a large number of different vacua. The potential contains a term of the form
\begin{equation}
V\sim {\rm Tr}\left(\frac{\mu}{3R}X^a +i\epsilon^{abc}X^bX^c\right)^2
\end{equation}
Classical vacua are
\begin{equation}
X^a(t)= \frac{\mu}{3R}J^a 
\end{equation}
where $J^a$ are matrices which satisfy the $SU(2)$ Lie algebra,
\begin{equation}
\left[J^a, J^b\right]=i \epsilon^{abc}J^c
\end{equation}
and could be either irreducible or reducible representations of $SU(2)$.
The representations of $SU(2)$ are interpreted as fuzzy spheres which are spherically symmetric states of membranes \cite{Berenstein:2002jq} 
(see also \cite{Myers:1999ps,Polchinski:2000uf}). 
They approach a classical membrane when the spin of an irreducible representation is large, the state with a single membrane being described by a single $N$-dimensional irreducible representation
of $SU(2)$ with spin $j=\frac{N-1}{2}$.

Recently, an alternative interpretation of these solutions as transverse five-branes has been given \cite{Maldacena:2002rb}.  In principle, any of the classical vacua corresponds to a five-brane state.  A single classical five-brane, with its five spatial dimensions forming a five-sphere and its time direction lying along the light-cone, is believed to correspond to the vacuum where the representation of $SU(2)$ is highly reducible, consisting of $N$ repetitions of the singlet $J^i=0$ representation. A detailed comparison of the fluctuation spectrum of the matrix model about these states with the fluctuations of the classical five-brane on a plane-wave background was given in ref.\ \cite{Maldacena:2002rb} and good agreement was found.  A stack of $k$ coincident five-branes is thought to correspond to the state with  $n=N/k$ repetitions of the $k$-dimensional representation of $SU(2)$. If the representations have differing dimensionalities, with maximum
dimension $k$, this still corresponds to $k$ spherical five-brane states where the branes have different radii.

The main difference between a five-brane state and a membrane state is that, for the five-brane, as $N$ is taken to infinity, the number of repetitions of
irreducible representations of $SU(2)$ is taken to infinity with their dimensionalities held fixed, whereas for the
membrane, the number or repetitions of an irreducible representation (number of membranes) is held fixed and the dimensionalities are taken to infinity.

The distinction between these two cases that will be important to us in the following occurs in the way that the gauge symmetry of (\ref{mmaction})
is realized.  Vacua with non-trivial representations of $SU(2)$
break the gauge symmetry, realizing it in a Higgs phase.  There is a  residual gauge symmetry which interchanges representations of the same dimensionality.  If there are $n_k$ representations of dimension $k$, the
gauge symmetry breaking pattern is
\begin{equation}
\frac{U(N)}{U(1)} \to \prod_{k=1}^\infty\left( \frac{U(n_k)}{U(1)}\right)
\label{pattern}\end{equation}
Here, because all degrees of freedom transform in the adjoint representation,
the true gauge group is the group modulo its center.
For a membrane state, as we take the limit $N\to\infty$,
the rank of the residual gauge group remains finite,
whereas for a five-brane, it becomes infinite.

The coupling constant which governs the size of corrections to
perturbation theory depends on the classical vacuum about which one is expanding.  For membrane vacua, the effective coupling constant is
\be
g_{\rm eff}=\left( \frac{3R}{\mu}\right)^{\frac{3}{2}}
\ee
For the five-brane, where the multiplicity of representations is $n$,
the effective coupling is the 't Hooft coupling
\begin{equation}
g_{\rm eff}^2 n=\lambda\equiv \left( \frac{3R}{\mu}\right)^3 n
\label{thooft}
\end{equation}
The difference comes from the residual $U(n)$ symmetry of the five-brane solution and the fact that
index loops in perturbation theory contribute factors of $n$ which can only
be controlled in the 't Hooft limit,
\begin{equation}
n\to\infty
~~~{\rm with}~~~
\lambda~{\rm fixed}
\label{thooft1}
\end{equation}

The decoupling limit, which isolates the theory living on
the five-brane, is the same as the 't Hooft limit in (\ref{thooft1}). Intuitively, this is the weakest coupling limit that can be taken that retains any of the interactions in the matrix model.  If it were taken for a membrane state, the matrix model would be non-interacting.  It is only the five-brane states which retain interactions.
The 't Hooft coupling is the radius of the five-brane squared in units of the
string scale
\be
\frac{r^2}{\alpha'} \sim \lambda
\ee

In this work we will be concerned with perturbative computations in five-brane 
backgrounds in the 't Hooft limit.
Our perturbative analysis is limited to small $\lambda$, the situation where
the five-brane is highly curved. In the end this will prevent us from making a 
direct comparison of our results with the properties of the usual 
NS five-brane of string theory which is large and flat and which we believe 
would be obtained by  the infinite radius, and therefore infinite 't Hooft 
coupling limit, in these sectors of the matrix model.

In this paper we shall analyze the
thermodynamics of this matrix model. The main observation will be that the five-brane states of the matrix model have a first order phase transition at some critical temperature.  This phase transition is seen in the 't Hooft limit.
Our analysis is further limited to the regime where the 't Hooft coupling
$\lambda$ is small.  However, the existence of the phase transition appears
robust and is expected to occur over a  range of $\lambda$.

We will show this in two different ways.  The first way counts the degeneracy of states with high energies in the infinite $N$ limit and establishes that there is a Hagedorn density of states.  From the exponential behavior we can find
the Hagedorn temperature.

Then, secondly, we use matrix model techniques to show that there is a
Gross-Witten type of large $N$ phase transition in the perturbative
limit of the model. In the weak coupling limit, the transition temperature coincides with the Hagedorn temperature that we extracted from the asymptotic density of states.  The advantage of the matrix model approach is that it is
at least in principle possible to study the phase above the Hagedorn transition.
The transition is similar to deconfinement transitions which are expected to occur in confining gauge theories at high temperature.

\vspace{12pt} \noindent
{\bf Note Added} \newline \noindent
While this research was conducted,
an extensive study of the Hagedorn transition 
in the case of weakly coupled $SU(N)$ 
gauge theories on compact spatial manifolds
was carried out
in an interesting work  \cite{AMvR:paper}. This work has a partial technical
overlap with ours.

\section{Hagedorn Behaviour of Matrix Models}
\label{HagedornMM}

In this section, we will compute the Hagedorn temperature of matrix models by 
counting their gauge invariant states, or, equivalently, the numbers of gauge 
invariant operators. We will show that, when the dimension $N$ of the matrices
is infinite, they have a Hagedorn density of states. 
Our analysis is very 
similar to that in ref.\ \cite{Sundborg:1999ue,Polyakov:2001af,AMvR:seminars}.

\subsection{A Toy Model}

We will begin with a toy model which captures the salient features of the
physics involved.  Consider a set of $d$ bosonic $N\times N$ Hermitian
matrices, $X_j(t)$, where $j=1,\ldots,d$ and $t$ is the time.
These are coupled to an $N\times N$ ``gauge field'' $A(t)$.  The action is
\begin{equation}\label{toyaction}
S=\int dt \sum_{j=1}^d\frac{1}{2}\left( (\frac{d}{dt}X_j+i[A,X_j])^2
-\omega^2 X_j^2\right)
\end{equation}
If it were not for the gauge field appearing in the covariant derivative
this would be a theory of $d$ independent matrix oscillators.
Introducing the gauge field makes the model invariant under the gauge transformation
\begin{eqnarray}
X_j(t)\rightarrow U(t)X_j(t)U^{\dagger}(t)\nonumber\\
A(t)\rightarrow U(t)A(t)U^{\dagger}(t)-iU(t)\dot U^{\dagger}(t)
\label{toygt}
\end{eqnarray}
The equation of motion which is obtained by varying $A(t)$
has no time derivatives and is an equation of constraint
\begin{equation}\label{gl}
{\cal G}=\sum_{j=1}^d\left[ X_j(t), \Pi_j(t)\right] \sim 0
\end{equation}
where
\begin{equation}
\Pi_j(t)=
\frac{d}{dt}X_j(t)+i\left[ A(t),X_j(t)\right]
\end{equation}
is the canonical momentum, with Poisson bracket
\begin{equation}
\left\{ X_j^{ab}(t),\Pi_k^{cd}(t)\right\} = \delta_{jk}\delta^{ad}\delta^{bc}
\end{equation}
The left-hand-side of the constraint equation (\ref{gl}) is the generator of 
an infinitesimal time-independent gauge transformation,
\begin{equation}
\delta X_j=
\left\{ \tr \left(\Lambda{\cal G}\right), X_j\right\}= -
\left[ \Lambda, X_j\right]
\label{tindgt}
\end{equation}
When we quantize, 
this toy model is simply described by $d$ decoupled matrix oscillators
with the constraint (\ref{gl}) on the physical states.
\footnote{
Obviously, in a deconfined phase this no longer holds for the vacuum as the
center symmetry is broken.}

\noindent
\begin{claim}  In the limit $N\to\infty$ this toy model has a Hagedorn density
of states and Hagedorn temperature
\begin{equation}
T_H=\omega/\ln d
\label{hagtemp}
\end{equation}
The case $d=1$ should be interpreted as $T_H = \infty$.  We will also use the parameter $\beta=1/T$ for the inverse temperature and the notation
$\beta_H=1/T_H=\ln d/\omega$.
\end{claim}

\noindent
To substantiate this claim, we will count the number of physical states which
have a given asymptotically large energy and extract $\beta_H$ from the exponential growth (if there is any).

The Hamiltonian which follows from canonical quantization of the action (\ref{toyaction}) is
\begin{equation}
H= \sum_{j=1}^d\frac{1}{2}{\rm Tr}\left( \Pi_j^2+\omega^2 X^2_j\right)
\label{toyham}
\end{equation}
where the momentum obeys the canonical commutation relation
\begin{equation}\label{toycom}
\left[ X_j^{ab}, \Pi_k^{cd}\right]=i\delta_{jk}\delta^{ad}\delta^{bc}
\end{equation}
It is easy to see that the constraint
(\ref{gl}) commutes with the Hamiltonian.

We shall take the strategy of quantizing the model of free oscillators with Hamiltonian (\ref{toyham}) and commutators (\ref{toycom}).  From the resulting quantum states we then isolate a subspace of physical states which are annihilated by the constraint operator in equation (\ref{gl}).

We define creation and annihilation operators
\begin{equation}
a_j=\frac{1}{\sqrt{2\omega}}\left( \Pi_j-i\omega X_j\right)
~~,~~
a_j^{\dagger}=\frac{1}{\sqrt{2\omega}}\left( \Pi_j+i\omega X_j\right)
\label{toycomm}
\end{equation}
so that
\begin{equation}
\left[ a_j^{ab}, {a_k^{\dagger}}^{cd}\right]
=\delta_{jk}\delta^{ad}\delta^{bc}
\end{equation}
and, after dropping the ground state energy, the Hamiltonian is
\begin{equation}
H=\sum_{j=1}^d ~\omega~\tr\left( a_j^{\dagger} a_j\right)
\end{equation}
The vacuum state is annihilated by all of the annihilation operators
\begin{equation}
a^{ab}_j \left| 0\right>=0~~~\forall ~a,b=1,...,N;j=1,...,d
\end{equation}
A basis of quantum states is found by operating creation operators
on the vacuum,
\begin{equation}
\left|0\right>
~~,~~
{a_j^{\dagger}}^{ab}\left| 0\right>
~~,~~
{a_j^{\dagger}}^{ab}{a_k^{\dagger}}^{cd}\left|0\right>
\ldots
\label{toybasis}
\end{equation}
These basis vectors are eigenstates of the Hamiltonian.

From these states, we must choose a subspace which is annihilated by the
constraint operator in (\ref{gl}).  That operator generates  infinitesimal
time independent gauge transformations.  Quantum states will be invariant if 
they are invariant under the finite transform
\begin{equation}
a_j\to U a_j U^{\dagger}
~~,~~
a_j^{\dagger} \to U a_j^{\dagger} U^{\dagger}
~~,~~
{\rm with}~U U^{\dagger} = 1
\label{fintimindgt}
\end{equation}
Such invariant states are generally obtained from the general basis states
by taking traces over the matrix indices,\footnote{If $N$ were finite there
would also be the possibility of other invariants such as the determinant of
$a_j^{\dagger}$.  The energy of the state created by this operator is $N\omega$.}
\begin{equation}
\left|0\right>
~~,~~
\tr\left(a_j^{\dagger}\right)\left| 0\right>
~~,~~
\tr\left(a_j^{\dagger}a_k^{\dagger}\right)\left|0\right>
~~,~~
\tr\left(a_j^{\dagger}\right)
\tr\left(a_k^{\dagger}\right)\left|0\right>
\ldots
\label{toybasisTrace}
\end{equation}

The operators which create gauge invariant states are thus of the form
$$
\tr \left(a_{j_1}^\dagger \ldots a_{j_n}^\dagger\right)
$$
and can be thought of as ``words'' which are made from the ``letters''
$a_i^{\dagger}$. The length of a word is the number of letters which it contains.
There are $d$ different letters at our disposal.  Inside a word, the ordering
of the letters is important, since they are matrices and do not commute with each other. However, because of the cyclicity of the trace, words which are related by cyclic permutations of the letters are equivalent.

Note that for finite $N$ the words are not independent.  Generally, a very long word can be expressed in terms of linear combinations of sentences made from shorter words: in the $U(N)$
theory, the operators $\tr (a^\dagger)^k$ for $k > N$ can be expressed in
terms of those with $0 \le k \le N$. However, the long words do become independent when the limit $N\to \infty$ is taken.  Here, we shall always assume that the large $N$ limit is to be taken.  For this reason, we shall treat all distinct words as independent operators.

Finally, we can create a state by applying several ``words'' to the vacuum.
Such multi--trace operators,
$$
\tr \left(a_{j_1}^\dagger \ldots a_{j_n}^\dagger\right)
\tr \left(a_{k_1}^\dagger \ldots a_{k_m}^\dagger\right)
\ldots
\tr \left(a_{\ell_1}^\dagger \ldots a_{\ell_p}^\dagger\right)
$$
will be called ``sentences''.  The
length of a sentence is the sum of the lengths of its constituent
words, or the total number of  letters in the sentence.
As each creation operator, or letter, creates
an excitation of energy $\omega$, the energy is the length of the sentence
in units of $\omega$.

Obviously, words or sentences should be considered equal if they produce the
same quantum state. As operators, words commute with each other.  Therefore, the order of the words in a sentence does not make any
difference.  

In order to find the Hagedorn temperature we should find the density of states
at asymptotically large energy.  If the theory has a Hagedorn spectrum,
the increase in the density of states will be exponential,
\be
\rho(E) \sim e^{\beta_H E}
\ee
This formula can be corrected by prefactors; we shall discuss them later, 
at subsection \ref{prefactor}.
We will concentrate for the time being on extracting the coefficient in the 
exponent which we shall identify with the inverse of the Hagedorn temperature.

To find the density of states, we are essentially looking for the number $s_n$ of
sentences of length $n$. In order to compute this number, we will need to consider
also the number $w_k$ of words of length $k$.
It is convenient to define 
\be
s_0 = w_0 = 1
\ee
As is common in combinatorial problems, it is helpful to define the generating
functions of those numbers,
\ba
W(x) & \equiv & \sum_{k = 0}^\infty w_k x^k \\
S(x) & \equiv & \sum_{n = 0}^\infty s_n x^n
\ea

\subsubsection{The $d = 1$ Case}

For the case $d = 1$ of an alphabet with a single letter $a^\dagger \equiv a_1^\dagger$, there is obviously only one word $\tr (a^\dagger)^k$ of each length $k$,
or $w_k = 1$. A sentence of length $n$ is nothing but a collection of such
words, where their order is immaterial. Hence, the number of sentences of
length $n$ is just the number of partitions of $n$ into natural numbers,
$s_n = p(n)$. It is well known from the days of Euler that
\begin{eqnarray}
\label{Px}
P(x) \equiv \sum_{n = 0}^\infty p(n) x^n &=&
(1 + x + x^2 + \ldots)(1 + x^2 + x^4 + \ldots)(1 + x^3 + x^6 + \ldots) \cdots
\nonumber\\&=&
\prod_{i=1}^\infty (1 - x^i)^{-1}
\end{eqnarray}
and from the celebrated work of Hardy and Ramanujan that
\be
\label{pn}
p(n) \sim e^{c \sqrt{n}}
\ee
with $c = \sqrt{\frac{2}{3}} \pi \approx 2.5651$.\footnote{
In fact, the result is much stronger. In particular \cite{Edwards},
$$
p(n) =
\frac{1}{4 \sqrt{3} \tn} \left(1 - \frac{1}{c \sqrt{\tn}} \right)
e^{c \sqrt{\tn}} +
O\left(\frac{1}{\tn} e^{c \sqrt{\tn} / 2}\right)
~,$$ where $$\tn = n - \frac{1}{24} ~.$$
}

The number of sentences of length $n$ is therefore sub-exponential, and the
partition function converges for all finite temperatures; consequently we
conclude that $\beta_H=0$ and in this case $T_H = \infty$ as claimed.

\subsubsection{The $d > 1$ Case}
\label{Mdl1}
An upper bound on the number of words of length $k$ is
$$w_k \le d^k$$ as there are $d$ independent choices for each letter.
The cyclic property of the trace makes some of those words equivalent.
Obviously, each word can have at most $k$ equivalent representations;
therefore, a lower bound on the number of inequivalent words is
$$w_k \ge \frac{1}{k} d^k$$ We claim that the latter bound is in fact a good
approximation, $$w_k \approx \frac{1}{k} d^k$$
To substantiate this claim, we will utilize the famous Polya theorem.
\begin{theorem}[Polya]
Let $G$ be a group acting on a set $X$. Then the number of colourings of the
elements of $X$ with $d$ colours, distinct under the action of $G$, is
\begin{equation}\frac{1}{|G|} \sum_{g \in G} d^{\lambda(g,X)}\end{equation} where $\lambda(g,X)$ is the
number of cycles in the action of $g$ on $X$.
\end{theorem}
In our case, $X = {\mathbb Z}_k$ is the string of letters in a word, and the
group of rotations is also $G = {\mathbb Z}_k$, acting on $X$ by addition.
The colours are the different kinds of letters, so their number is indeed $d$.
The number of colourings is just the number of inequivalent words.

Let us begin with the case when $k = p$ is a prime number. Then, the identity
has $|X| = p$ orbits, and each other rotation has only one; therefore,\footnote{
Incidentally, the fact that $w_p$ is integer provides a simple proof of Fermat's
little theorem, $d^p \equiv d \pmod{p}$.}
\be
w_p = \frac{1}{p} \left( d^p + (p-1) d \right)
\ee
This is indeed very close, for large $p$, to $\frac{1}{p} d^p$.
For the case where $k$ is not a prime, a similar formula can be written in
terms of Euler's totient function, but it suffices to note that
apart from the identity, again having $|X| = k$ orbits, all other rotations can
have at most $k/2$ orbits (in fact, there can be at most one such orbit, for
$g = k/2$ when $k$ is even). Therefore
$$w_k \le \frac{1}{k} \left( d^k + (k-1) d^{k/2} \right)$$ which is again close
to $\frac{1}{k} d^k$.

Accordingly, we shall henceforth use the approximation
$$w_k \approx \frac{1}{k} d^k$$ for $k \ge 1$.
This results in
\be
W(x) = \sum_{k = 0}^\infty w_k x^k \approx
1 + \sum_{k = 1}^\infty \frac{1}{k} d^k  x^k =
1 - \log (1 - d x)
\end{equation}

A sentence can be composed of only one word; The contribution to $s_n$ is then
$w_n$, and to $S(x)$ is $W(x)$. It can be composed of two words;
the contribution to $s_n$ is then naively the convolution
$(w \ast w)_n = \sum_{m = 0}^n w_m w_{n-m}$, and to $S(x)$ is $W(x)^2$, but we
should remember that the order of the words is immaterial. As for large $n$
the sentences composed of two identical words are negligible (the generating
function of those is $W(x^2)$), we should approximately divide by two.
In general we approximate that the number of words is so large that
generically no two will be identical in a sentence; if there are $a$ words in
a sentence, we should divide the naive approximation by $a!$. Consequently,
\begin{equation}
\label{Sx}
S(x) \approx \sum_{a = 0}^\infty \frac{1}{a!} W(x)^a =
e^{W(x)} \approx e \cdot \frac{1}{1 - d x} =
e \cdot \sum_{n = 0}^\infty d^n x^n
\end{equation}
so
\begin{equation}
s_n \approx e \cdot d^n \sim e^{(\log d) n}
\end{equation}
and the Hagedorn temperature, in units of $\omega$, is indeed
$\frac{1}{\log d}$ as claimed.

Note that if
\begin{equation}
\label{xbeta}
x = e^{- \beta \omega}
\end{equation}
then
\be
\label{ZequivS}
Z(\beta) \equiv S(x) = \sum_{n = 0}^\infty s_n e^{-\beta n \omega}
\ee
is exactly the partition function, so in light of (\ref{Sx}) we get
\begin{equation}
\label{Zbeta}
Z(\beta) \approx e \cdot \frac{1}{1 - d e^{-\beta \omega}}
\end{equation}
The partition function clearly diverges at the inverse Hagedorn temperature
$\beta_H$ such that
\begin{equation}
\label{betaHcond}
d \cdot e^{-\beta_H \omega} = 1
\end{equation}
which gives indeed $\beta_H = \frac{\log d}{\omega}$.

In fact, a more careful use of Polya's theorem can give an exact expression 
\cite{Sundborg:1999ue} for
$Z(\beta)$ or $S(x)$, but this is irrelevant for our 
purposes. 

\subsection{Extensions}

\subsubsection{The Traceless Case}
The matrices we have encountered thus far were assumed to be Hermitian.
They can equivalently be characterized as matrices in the adjoint
representation of the $U(N)$ algebra. It is natural to explore the variant
where the matrices are in the adjoint of $SU(N)$ (we will denote the various
magnitudes for the $SU(N)$ version with a tilde).
We will now show that our results are essentially unchanged if we have the
gauge group $SU(N)$ instead of $U(N)$.

A matrix in the adjoint representation of $SU(N)$ is Hermitian, but with the
extra constraint of being traceless. In particular $\tr a^\dagger = 0$ and
so there are no single--letter words:
\begin{equation}
\label{wtilde1}
\tilde{w}_1 = 0
\end{equation}

In the $d = 1$ case, therefore, the partitions of $n$ should not contain ones.
As any partition without ones of any $m \le n$ can be uniquely completed,
with the addition of $n-m$ ones, to a partition of $n$, and vice versa,
obviously 
\be
s_n = \sum_{m = 0}^n \tilde{s}_m
\ee
and so
\be
\tilde{s}_n = s_n - s_{n-1} \approx \frac{\partial}{\partial n} s_n \sim
\frac{1}{\sqrt{n}} s_n
\ee
We see that the exponential behaviour, and hence
the Hagedorn temperature, remains as in the $U(N)$ case, and the change is only
in the irrelevant polynomial prefactor. All this can also be seen from the
fact that the relevant partition function is
\be
\tilde{P}(x) = \prod_{i=2}^\infty (1 - x^i)^{-1} = (1 - x) P(x)
\ee

In the $d > 1$ case, (\ref{wtilde1}) implies
\be
\tilde{W}(x) = W(x) - d x
\ee 
and
\begin{eqnarray}
\tilde{S}(x) & \approx & e^{\tilde{W}(x)} = e^{-d x} e^{W(x)} \approx 
   e^{-d x} S(x) \nonumber \\
             & \approx &
   e \, (1 - d x + \frac{1}{2} d^2 x^2 - \frac{1}{3!} d^3 x^3 \pm \ldots)
   \, (1 + d x + d^2 x^2 + d^3 x^3 + \ldots)
\end{eqnarray}
so the coefficient of $x^n$ is now
$$
s_n \approx e \, d^n
(1 - 1 + \frac{1}{2} - \frac{1}{3!} + \ldots + (-1)^n \frac{1}{n!})
$$
but the last factor is approximately $e^{-1}$, so
\begin{equation}
s_n \approx d^n
\end{equation}
We get the same Hagedorn temperature as in the $U(N)$ case, the only change
being in an irrelevant constant factor.

\subsubsection{Fermions}

Fermionic matrix degrees of freedom behave quite similarly to bosonic ones of
the same mass. We can form the fermionic creation operators $b_i^\dagger$.
Each entry of such a matrix is a Grassmannian number which squares to zero;
however, there are so many of those, (i.e.\ $N^2$) that when forming words of
finite length, none vanishes, and more generally, the words remain independent.
However, each word squares to zero, so we can not use the same word more than
once in a sentence.

For the $d = 1$ case, this means that the partitions of $n$ should be to
distinct numbers. The number of such partitions, $q(n)$, behaves also as
\be
q(n) \sim e^{c' \sqrt{n}}
\ee
when now $c' = \sqrt{\frac{1}{3}} \pi$.
The corresponding generating function for the sentences is easily seen to be\footnote{
Incidentally, the number $q(n)$ of partitions of $n$ into distinct parts
is equal to the the number of partitions into odd parts; this is an amusing
exercise in generating functions, as formally
\begin{equation}\prod_{i=1}^\infty (1 + x^i) = \prod_{j=1}^\infty (1 - x^{2 j - 1})^{-1} .\end{equation}
}

\be
S(x) = Q(x) \equiv \sum_{n = 0}^\infty q(n) x^n =
\prod_{i=1}^\infty (1 + x^i)
\ee

For the $d > 1$ case, we should again take into account the restriction that
no two equal words appear inside a sentence. However, we have already argued
that such cases are rare and used the approximation that they are negligible,
when we argued that the ``$a$ words in a sentence'' term $W(x)^a$ should be
divided by $a!$ (so that $S(x)$ is the exponent of $W(x)$). We conclude that
the Hagedorn temperature of the $d > 1$ fermionic matrix model is equal to
its bosonic counterpart.

\subsubsection{Different Masses}

When there are several matrices with different masses, there is
obviously still a Hagedorn density of states, with the Hagedorn temperature
between two bounds: the Hagedorn temperature of the same number of matrices
but all with the same mass, which is either the minimal or maximal one.
However, it is quite easy to generalize (\ref{betaHcond}) to get an exact
result.

\begin{claim}
For a model with $d > 1$ matrices, bosonic or fermionic, of masses
$\omega_i > 0$,
$i = 1 \ldots d$, the inverse Hagedorn temperature $\beta_H$ is given by the
solution of
\begin{equation}
\label{betaHcondgeneral}
\sum_{i = 1}^d e^{-\beta_H \omega_i} = 1
\end{equation}
\end{claim}

To see that, we first pick a mass scale $\omega$ and write
\be
\label{nu}
\omega_i = \omega \nu_i
\ee
The idea of the proof in the equal masses case
carries over when we define
\begin{equation}
\label{Fx}
F(x) = \sum_{i = 1}^d x^{\nu_i}
\end{equation}
as a generalization of $d x$. We still have
\begin{equation}
W(x) \approx 1 + \sum_{k = 1}^\infty \frac{1}{k} F(x)^k = 1 - \log (1 - F(x))
\end{equation}
and
\begin{equation}
\label{SF}
S(x) \approx e^{W(x)} \approx e \cdot \frac{1}{1 - F(x)}
\end{equation}
Where now $s_n$ should be interpreted as the number of states with energy 
$n \omega$ and not as the number of sentences of length $n$; indeed, $n$ 
doesn't even have to be integer.
 
Clearly, $S(x)$ diverges at
\be
\label{xH}
x_H \equiv e^{-\beta_H \omega}
\ee
where
\be
\label{FxH}
F(x_H) = 1
\ee

Still using (\ref{xbeta}), we get
\begin{equation}
Z(\beta) \approx e \cdot \frac{1}{1 - \sum_{i = 1}^d e^{- \beta w_i}}
\end{equation}
which diverges at the claimed inverse Hagedorn temperature. As $F$ is a
monotone decreasing function of $\beta$, where $F = d > 1$ for $\beta = 0$
and $F = 0 < 1$ for $\beta = \infty$, there is exactly one root of equation
(\ref{betaHcondgeneral}).

If $\omega_i = 0$ for some $i$, then there is an infinite
number of zero modes, the Hagedorn temperature is formally zero,
and we see that the relation (\ref{betaHcondgeneral}) continues to hold
formally.

It is also easy to see that this reasoning, properly interpreted, does not rely
upon the number of matrices being finite.

\subsubsection{Absence of Prefactor}
\label{prefactor}
We would also like to refine our analysis and find any energy power prefactor
multiplying the leading exponential in the density of states. Stated otherwise,
assuming that the density of states behaves as 
\be
\label{PrefactorRho}
\rho(E) \sim C \, E^\alpha \, e^{\beta_H E}
\ee
we wish to find $\alpha$.
\begin{claim}
There is no power-like prefactor in matrix models, i.e.\ $\alpha = 0$.
\end{claim}  

In order to see this, assume (\ref{PrefactorRho}), and compute the 
partition function,
\be
Z(\beta) = \int \rho(E) \, e^{-\beta E} \, dE = 
\int_0^\infty C \, E^\alpha \, e^{-(\beta-\beta_H)E} \, dE = 
\frac{C \, \Gamma(\alpha+1)}{(\beta - \beta_H)^{\alpha+1}}
\ee
or equivalently by (\ref{xbeta},\ref{ZequivS},\ref{xH}),
\be
S(x) = \frac{C \, \Gamma(\alpha+1) \, \omega^{\alpha+1}}
            {(\log (x_H/x))^{\alpha+1}}
\ee
It is easy to see that
\be
\frac{S(x) \, S''(x)}{S'(x)^2} = \frac{\alpha+2 - \log(x_H/x)}{\alpha+1}
\ee
so that $\alpha$ can be extracted from the behaviour of the latter expression
at the divergence,
\be
\label{AlphaFromGuess}
\left.\frac{S(x) \, S''(x)}{S'(x)^2}\right|_{x = x_H} = 
\frac{\alpha+2}{\alpha+1}
\ee

On the other hand, it is immediate from (\ref{SF}) that
\be
\frac{S(x) \, S''(x)}{S'(x)^2} = \frac{2 F'(x)^2 - (F(x) - 1) F''(x)}{F'(x)^2}
\ee
and therefore, from (\ref{FxH}),
\be
\label{AlphaFromFx}
\left.\frac{S(x) \, S''(x)}{S'(x)^2}\right|_{x = x_H} = 2
\ee
Equating (\ref{AlphaFromGuess}) and (\ref{AlphaFromFx}) we get indeed 
that $\alpha = 0$ as claimed.

Essentially, this behaviour is dictated by the fact that $F(x)$, a monotonous 
function, has a non zero derivative at $x = x_H$.

\subsection{Hagedorn Behaviour in Free Second Quantized Theories}
\label{HagedornQFT}

Let us study a free quantum theory of harmonic oscillators of frequencies
$\omega_i > 0$. 
Their number may be finite or infinite, and they might be bosonic or fermionic.
The corresponding Hamiltonian, after discarding the zero--energy, is
$H = \sum_i \omega_i a_i^\dagger a_i$. The Fock space states of the second 
quantized theory (the ``sentences'') are generated directly by applying the 
creation operators (the ``letters'') on the vacuum: as the creation operators
are not matrices, and no gauge invariance condition is imposed, there is no 
need for ``words'' in this context. 
\begin{claim}
The inverse Hagedorn temperature $\beta_H$ in this model is given by the 
solution of
\be
\label{betaHcondquant}
\sum_i e^{-\beta_H \omega_i} = \infty
\ee  
More accurately, it is the maximal value of $\beta$ where the left hand side 
diverges. 
\end{claim}
Note that this implies that the Hagedorn temperature for a system with a 
finite number of oscillators is always infinite, so there is no Hagedorn 
transition (for that, it is essential that there are no zero modes 
$\omega_i = 0$). Even for systems with an infinite number of oscillators
the Hagedorn temperature might be infinite. It is indeed infinite in the 
sensible cases where the model comes from a field theory in a finite volume.

To understand the claim, we will begin with the example of $d$ bosonic 
oscillators of equal frequency $\omega$. As in the the corresponding matrix
case of subsection \ref{Mdl1}, $s_n$ counts the number of sentences of length
$n$ made out of $d$ letters. The order of the letters does not matter, as the 
creation operators commute, and therefore $s_n = \binom{n+d-1}{d-1}$ 
(think of arranging $n$ circles and $d-1$ dividers in a line). As the density
of states $s_n \approx \mbox{const} \cdot n^{d-1}$ behaves polynomially 
and not exponentially, there is no Hagedorn behaviour.

This can also be easily seen from the generating function:
\be
\label{SxBinom}
S(x) = \sum_{n = 0}^\infty  \binom{n+d-1}{d-1} x^n = \frac{1}{(1 - x)^d} 
      = \prod_{i = 1}^{d} (1 + x + x^2 + x^3 + \ldots)
\ee
The term $x^j$ in the $i$-th multiplicand represents $j$ excitations of the
$i$-th oscillator. 

The generalization to different frequencies is immediate: using 
(\ref{xbeta},\ref{nu}), a bosonic oscillator $a^\dagger$ contributes a 
factor of $\frac{1}{1 - x^\nu}$, while a fermionic one can be excited at most
once, as $(b^\dagger)^2 = 0$, and so contributes a factor of 
$1 + x^\nu$. Accordingly,
\be
\label{SxGeneralQuant}
S(x) = \prod_{\mbox{Bosons  }} \frac{1}{1 - x^{\nu_i}}    \cdot
       \prod_{\mbox{Fermions}} \left(1 + x^{\nu_i}\right) 
\ee  

Now, take the logarithm:
\be
\label{LogSxGeneralQuant}
\log S(x) = \sum_{\mbox{Bosons  }} -\log \left(1 - x^{\nu_i}\right) +
            \sum_{\mbox{Fermions}}  \log \left(1 + x^{\nu_i}\right) 
\ee  
If there is a subsequence of the $\nu_i$ going to zero, then $\log S(x)$ 
diverges for all $x > 0$, so by (\ref{xbeta}), $\beta_H = \infty$;
this is clearly true also from (\ref{betaHcondquant}). Assume, on the other 
hand, that $\nu^* \equiv \inf \nu_i > 0$. Then for $1 > x > 0$ we have
$x^{\nu_i} \le x^{\nu^*} < 1$, so there is some $\eta < 1$ such that 
$
(1 - \eta) x^{\nu_i} \le 
-\log \left(1 - x^{\nu_i}\right) , \log \left(1 + x^{\nu_i}\right) \le
(1 + \eta) x^{\nu_i}
$.
Consequently, (\ref{LogSxGeneralQuant}) and the right hand of 
(\ref{betaHcondquant}) converge or diverge together. This concludes the proof 
of the claim.

As an example, let us take a classical (non--relativistic) string tied in its
two ends and free to oscillate in one transverse dimension. 
The string's frequencies are harmonics of some natural frequency, 
$\omega_i = i \cdot \omega$ for $i = 1, 2, 3, \ldots$. The number of states 
of energy $n \cdot \omega$ is again $p(n)$, the number of partitions of $n$,
which is subexponential (\ref{pn}), so there is no Hagedorn behaviour. 
$S(x) = P(x)$ is given by (\ref{Px}) that has a radius of convergence one
around the origin. In this case, The right hand side of (\ref{betaHcondquant}),
$\sum_i e^{-\beta \omega_i} = \sum_i x^i = \frac{1}{1-x}$ has indeed the same 
radius of convergence. If the string lives in $D$ spacetime dimensions, there 
are $D-2$ transverse directions, so $S(x) = P(x)^{D-2}$, which still displays
no Hagedorn behaviour.

The lack of Hagedorn behaviour is generic in field theories. In the free
case in a finite volume, the number of oscillators grows like the phase space,
or as a power of the energy. The result (\ref{betaHcondquant}) implies 
that we should look at expressions of the form 
$\sum_{i = 0}^\infty i^d x^i \sim \frac{1}{(1 - x)^{d+1}}$,
 as was shown, in a different context, in (\ref{SxBinom}). The divergence is 
still at $x_H = 1$, corresponding to $\beta_H = 0$ or $T_H = \infty$.

\section{The BMN Matrix Model}

As mentioned in the introduction, this model arises from the study of a 
certain M--theory plane wave which is the Penrose limit of the 
$\mbox{AdS}_4 \times S^7$ or $\mbox{AdS}_7 \times S^4$ solutions. 
An interacting matrix model arises from the DLCQ description of this 
background when there are $N$ units of momentum on the light-like circle 
\cite{Berenstein:2002jq}.

There are $11 - 2 = 9$
coordinates apart from the light cone ones, which break into two groups:
three coordinates which are equivalent under an $SO(3)$ rotation,
and the other six, equivalent under $SO(6)$. Those coordinates give rise
to nine bosonic matrices: three of mass $\frac{1}{3} \omega$
for the appropriate $\omega$, and six of mass $\frac{1}{6} \omega$.
The model is supersymmetric and correspondingly has eight fermionic
matrices, all of mass $\frac{1}{4} \omega$.

The various classical solutions of the matrix model are given by vanishing
$SO(6)$ (and fermionic) matrices, and by the $SO(3)$ matrices being a
dimension $N$ representation of $SU(2)$.
Taking the small 't Hooft limit makes the theory free around the classical
solution.

\subsection{A Single Five-Brane}
A single spherical five-brane solution is represented by the classical
solution with the $SO(3)$ matrices vanishing (or being in the trivial
representation) \cite{Maldacena:2002rb}. This vacuum manifestly respects
the $U(N)$ gauge symmetry of the model. The oscillators are then just
given by the bosonic and fermionic $U(N)$ matrices described above.

According to our previous result (\ref{betaHcondgeneral}),
the inverse Hagedorn temperature is given by the solution of
\begin{equation}
3 e^{-\beta \omega / 3} + 8 e^{-\beta \omega / 4} + 6 e^{-\beta \omega / 6}
= 1
\end{equation}
or, if $\hat{x} =  e^{-\beta \omega / 12}$, by the solution
$0 \le \hat{x} \le 1$ of
\begin{equation}
(\hat{x} + 1)^3 (\hat{x} - \frac{1}{3}) =
3 \hat{x}^4 + 8 \hat{x}^3 + 6 \hat{x}^2 - 1 = 0
\end{equation}
that is, by $\hat{x} = \frac{1}{3}$, so
$\beta_H = - \log \hat{x} \cdot \frac{12}{\omega} =
\frac{12 \log 3}{\omega} \approx \frac{13.1833}{\omega}$
or $T_H \approx 0.0758533 \omega$.

\subsection{Multiple Five-Branes}
A solution of $k$ coincident spherical five-branes corresponds to the matrix
model classical vacuum with the $SO(3)$ matrices being $n$ copies
of the $k$ dimensional representation, where $n \equiv N/k$ is
taken to infinity, with $k$ fixed. The $SO(3)$ matrices having a vacuum
expectation value results in the Higgsing of the gauge group to
$U(n)$. Each $U(N)$ matrix is decomposed into $k^2$ different
$U(n)$ matrix oscillators, which arrange themselves into multiplets
of $SO(3)$ spins $j$, integer for the bosonic oscillators and half integer
for the fermionic ones.
As $k$ grows, more and more oscillators are added, as the maximal $j$
grows by one for each oscillator type. Essentially, each oscillator becomes
a tower of oscillators, with the original $k = 1$ oscillator at the bottom,
and two new towers begin at $k = 2$.
As a consequence, the Hagedorn temperature is lowered.
We will see, however, that the change is very small even in the large $k$
limit, because the masses of the oscillators grow linearly in $j$
(their degeneracy does also, of course, but this is much less important).

The spectrum of oscillators in this case was computed in
\cite{Dasgupta:2002hx,Dasgupta:2002ru}, and is given (in units of $\omega$) in 
table \ref{TableFivebranes}, where we also give the labels given there to 
the different oscillator towers.
\begin{table}[ht]
\begin{center}
\begin{tabular}{|c||c|c|c|c|c|}
\hline
{\rm Type} & {\rm Label} & {\rm Mass} & {\rm Spins} & $SO(6) \times SO(3)$
\\ \hline\hline
{\rm Bosonic} $SO(6)$ & $x     $ & $\frac{1}{6 } + \frac{j}{3}$ &
$0 \le j \le k-1              $ & $(6       ,2j+1)$
\\ \hline
{\rm Bosonic} $SO(3)$ & $\alpha$ & $\frac{1}{3 } + \frac{j}{3}$ &
$0 \le j \le k-2              $ & $(1       ,2j+1)$
\\
& $\beta $ & $               \frac{j}{3}$ &
$1 \le j \le k                $ &  $(1      ,2j+1)$
\\ \hline
{\rm Fermionic}       & $\chi  $ & $\frac{1}{4 } + \frac{j}{3}$ &
$\half \le j \le k-\frac{3}{2}$ & $(\bar{4} ,2j+1)$ \\
& $\eta  $ & $\frac{1}{12} + \frac{j}{3}$ &
$\half \le j \le k-\half      $ & $(4       ,2j+1)$
\\
\hline
\end{tabular}
\caption{Oscillator masses for $k$ coincident five-branes
(adapted from \cite{Dasgupta:2002hx})}
\label{TableFivebranes}
\end{center}
\end{table}

The contribution of the oscillators labeled by ``$x$'' is
\begin{equation}
\sum_{j = 0}^{k-1} 6(2j+1) \hat{x}^{4j+2} =
\frac{6\left( \hat{x}^2 + \hat{x}^6 -
(2k+1) \hat{x}^{4k+2} + (2k-1) \hat{x}^{4k+6} \right)}
{(1 - \hat{x}^4)^2}
\end{equation}
which converges in the large $k$ limit (as $\hat{x} < 1$) to
\begin{equation}
\frac{6\left( \hat{x}^2 + \hat{x}^6 \right)}{(1 - \hat{x}^4)^2} =
\frac{6 \hat{x}^2 (1 + \hat{x}^4)}{(1 - \hat{x}^4)^2}
\end{equation}
The computations for the other kinds of oscillators are similar, and we
will only display the large $k$ limit result. The ``$\alpha$'' oscillators
contribute $\hat{x}^4 (1 + \hat{x}^4) / (1 - \hat{x}^4)^2$, the
``$\beta$'' oscillators $(3 \hat{x}^4 - \hat{x}^8) / (1 - \hat{x}^4)^2$,
the ``$\chi$'' oscillators $8 \hat{x}^5 / (1 - \hat{x}^4)^2$
and the ``$\eta$'' oscillators $8 \hat{x}^3 / (1 - \hat{x}^4)^2$.
The sum of all those contributions should equal one, so after multiplying
by $(1 - \hat{x}^4)^2$ we get
\begin{equation}
(\hat{x} + 1)^4 (\hat{x}^4 - 4 \hat{x}^3 + 4 \hat{x}^2 - 4\hat{x} +1) = 0
\end{equation}
having a unique real root between zero and one,
\begin{equation}
\hat{x} = 1 + \frac{1}{\sqrt{2}} - \sqrt{\sqrt{2} + \frac{1}{2}}
\approx 0.323556
\end{equation}
so
$\beta_H =       - \log \hat{x} \cdot \frac{12}{\omega}
\approx \frac{13.5406}{\omega}$ and
$T_H \approx 0.0738519 \omega$. Note how close these values are to those
of the single five-brane ($k = 1$) case.

\subsection{Membranes}
The matrix model vacuum in which the $SO(3)$ matrices are in the $N$ 
dimensional irreducible representation corresponds to a spherical membrane.
The vacuum with $k$ copies of the
$n$ dimensional irreducible representation, with 
$n \equiv N/k \rightarrow \infty$, and $k$ fixed, corresponds to $k$
coincident spherical membranes \cite{Berenstein:2002jq,Dasgupta:2002hx}.
The spectrum of excitations above those vacua was computed in 
\cite{Dasgupta:2002hx,Dasgupta:2002ru} and is given in table 
\ref{TableMembranes} (again in units of $\omega$).
\begin{table}[ht]
\begin{center}
\begin{tabular}{|c||c|c|c|c|c|}
\hline
{\rm Type} & {\rm Label} & {\rm Mass} & {\rm Spins} & $SO(6) \times SO(3)$
\\ \hline\hline 
{\rm Bosonic} $SO(6)$ & $x     $ & $\frac{1}{6 } + \frac{j}{3}$ & 
$0 \le j \le n-1              $ & $(6       ,2j+1)$ 
\\ \hline 
{\rm Bosonic} $SO(3)$ & $\alpha$ & $\frac{1}{3 } + \frac{j}{3}$ & 
$0 \le j \le n-2              $ & $(1       ,2j+1)$ 
\\ 
                      & $\beta $ & $               \frac{j}{3}$ & 
$1 \le j \le n                $ & $(1      ,2j+1)$  
\\ \hline
{\rm Fermionic}       & $\chi  $ & $\frac{1}{4 } + \frac{j}{3}$ & 
$\half \le j \le n-\frac{3}{2}$ & $(\bar{4} ,2j+1)$ \\  
                      & $\eta  $ & $\frac{1}{12} + \frac{j}{3}$ & 
$\half \le j \le n-\half      $ & $(4       ,2j+1)$ 
\\
\hline
\end{tabular} 
\caption{Oscillator masses for $k$ coincident membranes 
(adapted from \cite{Dasgupta:2002hx})}
\label{TableMembranes}
\end{center}
\end{table}

In the single membrane case, the gauge group is completely Higgsed, and the 
oscillators are not matrices anymore. Although there is an infinite number
of such oscillators as $N \rightarrow \infty$, their density is just linear
(the mass grows as the spin $j$ while the degeneracy is proportional to 
$2j + 1$), and as explained in subsection \ref{HagedornQFT}, the Hagedorn 
temperature is infinite.

In the case of $k$ coincident membranes, the gauge group is Higgsed down to 
$U(k)$, and the oscillators are $k$ dimensional matrices. Each oscillator, 
then, is comprised of $k^2$ entries, a fixed finite number, and viewing
those as the basic oscillators, even disregarding the gauge invariance 
condition, shows that the density of oscillator masses is still insufficient
to produce a Hagedorn behaviour.

\section{Thermodynamics of the Matrix Model}

In the previous section, we demonstrated that the spectrum of the matrix
model has a Hagedorn density of states in the large $N$ limit and that
we should expect a phase transition at some temperature of order the
parameter $\omega$.  In this Section, we will demonstrate that this is indeed
the case using a direct analysis of the thermodynamics of the matrix model.
We will be interested in the thermodynamic partition function, where we take the Hamiltonian, $H$, for the model corresponding to (\ref{mmaction}) and
form the partition function
$$
Z[N,R,\beta]~=~\tr e^{-\beta H} ~=~ e^{-\beta F[N,R,\beta]}
$$
where $\beta=1/T$ is the inverse temperature and
$F[N,R,\beta]$ is the free energy.
Before we discuss this, we shall again illustrate the method by using 
a simple toy model.

\subsection{The Toy Model}

Consider again the toy model of section \ref{HagedornMM}.
The thermodynamic partition function has a functional integral representation
\begin{equation}
Z=\int [dA][dX_j]e^{-\int_0^\beta d\tau\frac{1}{2}
\sum_{j}\tr\left( (DX_j)^2+\omega^2X_j^2\right)}
\label{toypart}
\end{equation}
where the fields have periodic boundary conditions in Euclidean time
\begin{equation}
A(\tau+\beta)=A(\tau)
~~,~~
X_j(\tau+\beta)=X_j(\tau)
\end{equation}
To proceed, we must fix a gauge.  The periodic boundary condition prevents
us from choosing the $A=0$ gauge.
The best we can do is to fix a gauge where the gauge field is static and diagonal,
\begin{equation}
\frac{d}{d\tau}A(\tau)=0
~~,~~
A^{ab}=\delta^{ab}A_a
\end{equation}
In that case,
\begin{equation}
(DX_j)^{ab}= \frac{d}{d\tau}X_j^{ab}+i(A_a-A_b)X_j^{ab}
\end{equation}
The Faddeev-Popov determinant for fixing this gauge is computed in the
Appendix.  It turns out to be similar to the Vandermonde determinant which
appears in the measure of unitary matrix integrals,

\begin{equation}
{\det} D~{\det}'\frac{d}{d\tau}~=~
\prod_{a\neq b}\sin\frac{\beta}{2}\vert A_a-A_b\vert
\end{equation}
where $\det'$ 
is a determinant restricted 
to the space of non-constant modes.
Then, the integral over the oscillator fields in (\ref{toypart}) is Gaussian
and can be performed explicitly

\begin{equation}
Z=\int \prod_{a=1}^N dA_a \prod_{a\neq b}\sin\frac{\beta}{2}\vert A_a-A_b\vert
\prod_{a,b}
{\det}^{-\frac{d}{2}}\left( -\left(\frac{d}{d\tau}+i(A_a-A_b)\right)^2
+\omega^2\right)
\label{202}
\end{equation}
The determinant can be evaluated explicitly (see the Appendix, also
ref.\ \cite{Ambjorn:1998zt}). This model with $d=1$, and a different critical behavior from the one which we shall be discussing in the following, has been studied extensively
in the context of lower dimensional string theories.  For example, see
refs.\ \cite{Boulatov:1991fp},\cite{Douglas:2003up}.

We obtain
\begin{equation}
Z=\int \prod_{a=1}^N dA_a \prod_{a\neq b}\sin\frac{\beta}{2}\vert A_a-A_b\vert
\prod_{a,b}\left(\frac{1}{ \sinh\left( \frac{\beta}{2}\left(\omega+i
(A_a-A_b)\right)\right) } \right)^d
\end{equation}
The effective action for the eigenvalues is then

\begin{equation}
S_{\rm eff}=\sum_{a\neq b}
\left\{
d\ln\left( \sinh\frac{\beta}{2}\left(\omega+i(A_a-A_b)\right) \right)-
\ln \sin \frac{\beta}{2} \left| A_a-A_b \right|
\right\}
\end{equation}
We have dropped an $A$-independent ground state energy.  There
are now $N$ remaining degrees of freedom, the variables $A_a$,
and because there are two $N$-fold summations, the action is of order $N^2$.
For this reason, in the large $N$ limit the integration in (\ref{202}) can be done in the saddle point approximation.  In that case there is a well-defined expansion
of the integral as a power series in $1/N^2$.   To begin, we must find
the saddle point by minimizing the effective action.
The equation for the minimum, for each $a$, is
\begin{equation}
\sum_{b\neq a} d\frac{ \sin\beta(A_a-A_b)}{\cosh\beta\omega-\cos\beta(A_a-A_b)}= \sum_{b\neq a}\cot\frac{\beta}{2}\left(A_a-A_b\right)
\end{equation}

It is useful to introduce the eigenvalue density,
\begin{equation}
\rho(\theta)=\frac{1}{N}\sum_{a=1}^N \delta(\theta-\beta A_a)
\label{density}
\end{equation}
and to assume it is a smooth function in the large $N$ limit.
Without loss of generality, $-\pi \le \theta \equiv \beta A \le \pi$,
and the density is normalized so that
\begin{equation}\label{norm}
\int_{-\pi}^\pi d\theta \rho(\theta)=1
\end{equation}
Obviously, the density is constrained to be positive, $\rho(\theta) \ge 0$.
Now, the action can be written as
\begin{equation}
\label{SeffRho}
S_{\rm eff}=N^2\int d\theta d\theta'
\rho(\theta)\rho(\theta')
\left\{
d \ln \left( \sinh \left( \frac{\beta \omega}{2} + \frac{i}{2}(\theta-\theta')\right)\right)-
\ln\left(\sin\frac{1}{2}\vert \theta-\theta'\vert\right)
\right\}
\end{equation}
We see there is a sort of competition between the eigenvalue repulsion of
the second term, coming from the Vandermonde, and the eigenvalue attraction
of the first term, coming from the action. The attraction is more pronounced
in the high temperature regime, $\beta \rightarrow 0$. 

The equation for the eigenvalue density is given by
\begin{equation}\label{speqn}
\int d\theta'\rho(\theta') d\frac{ \sin(\theta-\theta')}{\cosh\beta\omega-\cos(\theta-\theta')}= \int d\theta'\rho(\theta')\cot\frac{1}{2}\left(\theta - \theta'\right)
\end{equation}
where it is understood that the principal value around the singularity should
be taken in the right hand side integral.
Note that this equation should be satisfied only in the support of 
$\rho(\theta)$, that is, for angles $\theta$ such that $\rho(\theta) > 0$. 
This equation should be solved together with the normalization condition
(\ref{norm}) and the positivity constraint.

First, we observe that the constant eigenvalue density
\begin{equation}
\rho_0=\frac{1}{2\pi}
\end{equation}
is always a solution of (\ref{speqn}) where both sides of the equation vanish.
However, this solution is stable only for a range of values of $\beta$.  To see this, we Fourier expand inside (\ref{SeffRho}),\footnote{Here, we use the identity
$$
\ln\left(\sin\frac{\theta}{2}\right)
=-\sum_{n=1}^\infty \frac{\cos n\theta}{n}+{\rm const.}
$$
}
\begin{equation}
\label{SeffSumn}
S_{\rm eff}=N^2\int d\theta d\theta' \rho(\theta)\rho(\theta')
\sum_{n=1}^\infty \frac{1 - d e^{-n\beta\omega}}{n}\cos(n(\theta-\theta'))
\end{equation}
From this, we can see that if $\beta$ is large enough, then the phase where
the density is constant (so that the Fourier modes vanish,
$\int d\theta \rho(\theta)e^{in\theta} = 0$ for $n \neq 0$)
is stable.  As the temperature is raised, and $\beta$ is decreased, the
first instability sets in at
$\beta = \beta_H = 1/T_H$ where
\begin{equation}
T_H= \frac{\omega}{\ln d}
\end{equation}
This coincides with the Hagedorn temperature in (\ref{hagtemp})
which we found by estimating the high energy density of states in 
section \ref{HagedornMM}. 

Note that the approximations of section \ref{HagedornMM} amount
to keeping the $n = 1$ term only in (\ref{SeffSumn}), and that the arguments
there are invalid in the deconfined phase above the Hagedorn temperature. 
However, the result of section \ref{HagedornMM} correctly gives the free 
energy below the Hagedorn temperature. All that (\ref{SeffSumn}) 
tells us about the low temperature phase is that the free energy there 
is zero in terms of $N^2$; 
indeed, it is $O(N^0)$ in the confined phase, but more elaborate 
computations are needed in order to reproduce this from the matrix theory 
perspective. 

The straightforward generalization of this model where we have different frequencies for the bosonic oscillators and possibly some fermionic oscillators gives  the general equation for the critical temperature of the matrix model,
\begin{equation}
\sum_i e^{-\beta_H\omega_i}= 1
\end{equation}
This also coincides with the equation for the Hagedorn temperature which was
found in section \ref{HagedornMM}.

The phase transition that we have identified is the usual Gross-Witten \cite{Gross:he} transition of unitary matrix models.  In that phase transition the distribution of eigenvalues of the unitary matrix rearranges itself from one which has compact support near the identity matrix to one which is distributed
on the whole unit circle.

Above the phase transition the eigenvalue distribution is not uniform. 
Without loss of generality, we may assume it is an even function concentrated 
around $\theta = 0$. 
The distribution can be approximated by the semi-circle one,
\begin{equation}
\rho(\theta)=\frac{2}{\pi\xi}\cos\frac{\theta}{2}\sqrt{ \frac{\xi}{2}
-\sin^2\frac{\theta}{2}}
\label{semicircle}
\end{equation}
This distribution has support in the region $ \vert
\sin\frac{\theta}{2}\vert
\leq\sqrt{\frac{\xi}{2}}$ and the parameter $\xi$ is
restricted to the range $\xi<2$.
This approximation is exact when we truncate the low temperature expansion
of the matrix contribution to the effective action at its first nontrivial,
$\rho(\theta)$ dependent, term
\be
\label{SeffApprox}
d \ln (1 - e^{-\beta\omega - i(\theta-\theta')}) \approx 
-d e^{-\beta\omega} e^{-i(\theta-\theta')}
\ee
and the saddle point equation is approximately
\be
\int d\theta' \rho(\theta') 
     \left(2 d e^{-\beta \omega} \sin(\theta-\theta') - 
           \cot \frac{1}{2}(\theta-\theta')
     \right) = 0
\ee
This equation is solved by (\ref{semicircle}) when
\footnote{To see this, we need the integrals
$$
\int d\theta' \rho(\theta') \cot\frac{1}{2}(\theta-\theta')
=\frac{2}{\xi}\sin\theta
$$
where again the principal value is implied, and
$$
\int d\theta\rho(\theta) \cos\theta= \left(1-\frac{\xi}{4}\right)
$$
}
\be
\xi=2\left(1-\sqrt{1-1/de^{-\beta\omega}}\right)
\ee
This solution exists only when $de^{-\beta\omega}\geq 1$, that is
when the temperature is greater than the Hagedorn temperature. In this
region $\xi<2$ and the semi-circle distribution is well-defined.
Note, however, that the expansion parameter in the approximation 
(\ref{SeffApprox}), $e^{- \beta \omega}$, is at least 
$e^{- \beta_H \omega} = \frac{1}{d}$.

This approximation is not too bad, however, even in the high temperature 
limit. For $\beta = 0$ we get a narrow distribution, as 
$\xi \sim \frac{1}{d}$. It is easy, though, to solve exactly for the
density in this limit. Equation (\ref{speqn}) reduces, for $d > 1$, to
$$
\int d\theta' \rho(\theta') \cot\frac{1}{2}\left(\theta - \theta'\right) = 0
$$ 
Because of the implicit principle value we immediately see that in this limit,
$\rho(\theta) = \delta(\theta)$ is a solution (with support only at 
$\theta = 0$); the eigenvalue attraction totally wins over.
 
\subsection{Confinement-Deconfinement Transition}

The matrix model phase transition that we have found is characteristic
of a deconfining phase transition in a gauge theory which is heated to
a sufficiently high temperature.   The matrix quantum mechanics has the gauge
symmetry (\ref{toygt}) where, since all variables transform under the adjoint
representation, the unitary matrices are periodic up to an element of the center of the group,
\begin{equation}
U(\tau+\beta)=zU(\tau)
\label{centergt}
\end{equation}
and where $z$ is an element of the center of the group.  If the gauge group is
$U(N)$, the center is $U(1)$.  If it is $SU(N)$, the
center is $Z_N$, the multiplicative group of the $N$'th roots of unity.

No local operators are sensitive to the center element in (\ref{centergt}).
However, there is a nonlocal operator which transforms under the center,
the Polyakov loop operator
\begin{equation}
P=\tr {\cal P}\exp\left( i\int_0^\beta d\tau A(\tau)\right)
\label{ploop}
\end{equation}
where ${\cal P}$ denotes path ordering.
Under a gauge transform with boundary condition (\ref{centergt}),
\begin{equation}
P \to zP
\end{equation}

The center symmetry is often treated as a global symmetry of the finite
temperature gauge theory. As a global symmetry one can ask the question
as to whether or not it is spontaneously broken.  An order parameter
is the expectation value of the Polyakov loop, $\left< P \right>$.  
If the center symmetry is unbroken, this expectation value must vanish.  
If it is broken, then $\left< P \right>$ can be non-zero.

The vanishing or non-vanishing of $\left< P \right>$ has an interpretation 
in terms of confinement. The expectation value, 
\begin{equation}
\left< P \right> = 
\frac{ \int [dA][dX]e^{-\int_0^\beta d\tau\frac{1}{2}\tr((DX)^2 + \omega^2X^2)}
\tr{\cal P}e^{i\int_0^\beta d\tau A(\tau)}   }
{ \int [dA][dX]e^{-\int_0^\beta d\tau\frac{1}{2}\tr( (DX)^2+
\omega^2X^2)}   }
\equiv \exp\left( -\beta (F-F_0) \right)
\end{equation}
is the ratio of partition functions for the system where one classical
quark source is inserted to the partition function in the absence of the
source. $F-F_0$ is interpreted as the difference of the free energies
with and without the quark.  When the expectation value vanishes, this is interpreted as taking an infinite amount of energy to insert the quark.
Thus, the phase with unbroken center symmetry is confining.  When the
symmetry is broken, the free energy of the extra quark is finite and
that phase is deconfined.
Some examples of lower dimensional gauge theories where the expectation value can be computed 
appear in ref. \cite{Grignani:1995hx,Semenoff:1996ew,Semenoff:1996xg}. 
In particular, the case of the non-abelian coulomb gas could be written as an effective unitary 
matrix model and it exhibited a phase transition very similar to the one that we have studied here 
\cite{Semenoff:1996ew,Semenoff:1996xg}.

\subsection{The Matrix Model}

The variables in the action of the matrix model (\ref{mmaction}) can be
rescaled so that it has the form
\begin{eqnarray}
 \label{rescaled}
S&=&
\left(\frac{\mu}{3R}\right)^3
\int_0^{\beta\mu/3}
\!\!\!\!\!\!\!\!d\tau \,\, 
{\rm Tr}
\left( \frac{1}{2} (DX^i)^2
+ \psi^{\dagger}D\psi
+ \frac{1}{2} X^{a2}+\frac{1}{8} X^{a'2}
 + \frac{3}{4}
\psi^{\dagger}\psi 
\right. \nonumber \\
& &\qquad  \qquad \qquad \qquad \qquad \quad
+
\left.
i \epsilon^{abc}X^aX^bX^c
 +\psi^{\dagger}\sigma^a[X^a,\psi]
\right. \nonumber \\
& &-
\left.
\frac{1}{2}\epsilon_{\alpha\beta}\psi^{\dagger\alpha I}
g_{IJ}^{a'}
[X^{a'},\psi^{\dagger\beta J}]
+\frac{1}{2}\epsilon_{\alpha\beta}\psi^{\alpha I}
g_{IJ}^{a'\dagger}
[X^{a'},\psi^{\beta J}] 
- \frac{1}{4}[X^i,X^j]^2
\right)
\end{eqnarray}
Here, $\sigma^a$ are usual Pauli matrices and
$g_{IJ}^{a'}$ relate the $SU(4)$ fundamentals to
an $SO(6)$ fundamental \cite{Dasgupta:2002hx}.
As discussed in the Appendix, rescaling of the
coordinates does not affect the path integral measure.

We will consider the semiclassical expansion of this model about the
classical vacuum corresponding to the single five-brane,
$$
X_{\rm cl}=0
$$

Expansion to one loop order around this solution gives the semi-classical
partition function
\begin{equation}
Z\approx \prod_{a=1}^N
\int_0^{2\pi} [dA_a]\frac{   \left(
\det_F(D+3/4)\right)^8\vert{\det}_B(D)\vert\vert{\det}_B'(d/d\tau)\vert }
{ \left({\det}_B(-D^2+1)\right)^{3/2}\left( {\det}_B(-D^2+1/4)\right)^3 }
\label{semiclass}
\end{equation}
Where the differential operators in ${\det}_B$ and ${\det}_F$ have periodic and
anti-periodic boundary conditions, respectively.
We discuss how to evaluate the determinants explicitly in the Appendix.
The result is
\begin{equation}
Z\approx\prod_{a=1}^N
\int_0^{2\pi} [dA_a]\prod_{a\neq b}
\frac{
\left|\cosh \frac{1}{2} \left(\frac{\mu}{4T}+i(A_a-A_b)\right)\right|^8
\sin    \frac{1}{2} \left|                 A_a-A_b \right|
}
{\left|
\sin\frac{1}{2} \left(   \frac{\mu}{3T}+i(A_a-A_b) \right)\right|^{3}
\left|
\sin  \frac{1}{2} \left(   \frac{\mu}{6T}+i(A_a-A_b) \right)
\right|^6
}
\end{equation}

When $N$ is finite, this integral is a smooth function of $\mu/T$.
When $N$ is put to infinity, it can have singularities and it typically
has a phase transition at some critical value of $\mu/T$.  In that limit,
the action, which is the logarithm of the integrand, is of order $N^2$, whereas
there are $N$ degrees of freedom, $A_a$.  In the infinite $N$ limit, the
integral can be evaluated by saddle point technique.  When $\mu/T$ is a large number, the integrand is dominated by the Faddeev-Popov term which tends to spread the eigenvalues apart, giving a uniform distribution of $e^{iA_a}$ around the unit circle in the complex plane.  When $\mu/T$ is small, on the other hand, the
terms in the denominator cancel the Faddeev-Popov term and the
remainder of the integrand emphasizes configurations where the eigenvalues are
close together.  Because of the translation invariance of the potential, which
is a result of the center symmetry of the action, the accumulation point of the
eigenvalues is arbitrary and breaks the $Z_N$ symmetry of the effective theory.

In order to evaluate the integral in the infinite $n$ limit, we must find
the minimum of the action
$$
S_{\rm eff}=\int d\theta \rho(\theta)d\theta'\rho(\theta')\sum_{n=1}^\infty
\frac{1}{n}\left[1+8(-1)^ne^{-\mu n/4T}-3e^{-\mu n/3T}-6e^{-\mu n/6T}\right]
e^{in(\theta-\theta')}
$$
It is easy to see that, if the temperature is large enough, the potential
favors a non-zero value of the Fourier transforms
$$
\rho_n = \int_{-\pi}^{\pi} d\theta \rho(\theta) e^{i n \theta}
$$
In fact, the first one to condense is $\rho_1$ ($\rho_0=1$ is constrained 
by normalization).  What keeps it finite is the constraint that 
$\rho(\theta) \geq 0$ for all values of $\theta$.

The critical temperature for $\rho_n$ satisfies the equation
\begin{equation}
1+8(-1)^ne^{-\mu n/4T_{\rm crit}}-3e^{-\mu n/3T_{\rm crit}}-6e^{-\mu n/6T_{\rm crit}}=0
\label{trans}
\end{equation}

At the phase transition the free energy changes from a function which is
zero in leading order to one which is non-zero.  The phase transition is
first order.

The reason why one can have a phase transition in such a low dimensional system
(the only dimension is the one labeled by $\theta$ which comes
from the matrix index) is because the effective action is non-local, having
infinite range interactions.

\section{Discussion}

\subsection{M-Theory Perspective}

The energy scale $\omega$ of the excitations above the different vacua was
recognized in \cite{Dasgupta:2002hx} to be the scale $\mu$ appearing in the
M-theory plane wave background and in the matrix action (\ref{mmaction}).
The higher order interactions are governed by the coupling 
\be
g_{\rm eff} = \left( \frac{3R}{\mu}\right)^{\frac{3}{2}}
\ee
where we work in Planck length units, $\ell_p = 1$.
Moreover, in the five-brane case, the energy shifts of the excitation masses
from the free, quadratic, values are given by the 't Hooft coupling 
\be
\lambda \equiv g_{\rm eff}^2 N = \left(\frac{3R}{\mu}\right)^3 N
\ee 
so our results are valid only when $\lambda$ is small, $\lambda \ll 1$. 
Obviously we need also to take $N \rightarrow \infty$ for the 't Hooft limit 
in the matrix model. 
Furthermore, we need to keep $\mu$ fixed in order to get a nontrivial Hagedorn
temperature result. All this means that we take the limit in which the DLCQ 
radius vanishes, $R \rightarrow 0$. The brane light cone momentum, 
\be
\label{pplus}
p^+ = \frac{N}{R}
\ee   
therefore diverges in this limit, $p^+ \rightarrow \infty$.
The radius of the five-sphere on which the five-brane wraps can be computed 
in the small coupling regime \cite{Maldacena:2002rb}. Classically the $SO(6)$
directions are zero, but quantum mechanically, at one loop order, they give
\be
\label{r5}
r_5 = \left\langle \frac{1}{N} \tr \left(X^{a'}\right)^2 \right\rangle^{1/2} =
      \left( \frac{18 N^2}{\mu p^+} \right)^{1/2} = 
      \left( \frac{18 N R}{\mu    } \right)^{1/2} =
      \left( 6 \lambda^{1/3} N^{2/3}\right)^{1/2}
\ee 
so it diverges if $\lambda$ is kept fixed, but not necessarily if,
additionally, $\lambda$ is taken to zero in this limit.

The five-branes we have studied are therefore quite peculiar
from the M-theory point of view. 
As solutions with a light-like time direction, they are in the infinite 
momentum frame, moving (almost) at the speed of light.
They have large momentum on another light-like direction which is compact and 
small. This makes it difficult to compare them directly with ordinary 
NS five-branes or
M five-branes. The five-branes might be flat (with $r_5$ large in Planck 
length units) or not, according to the exact limit taken.

We would have liked, of course, to explore the behaviour of the five-branes 
in the large $N$ limit where $\mu$ and $p^+$ (and therefore $r_5$) are fixed. 
This however implies $R \rightarrow \infty$, 
so $g_{\rm eff} \rightarrow \infty$ and, a fortiori, 
$\lambda \rightarrow \infty$, where our approximations are invalid.

Anticipating the results of the next subsection, let us look at the 
five-sphere radius in terms of the length scale $\frac{\mu}{R}$ 
(which is the typical (classical) length scale of
our matrix model, as can be seen from the rescaling done in (\ref{rescaled})). 
We obtain
\be
r_5^2 \cdot \left(\frac{\mu}{R}\right)^{-2} = \frac{2}{3} \lambda
\ee
depends only on the 't Hooft coupling and is directly proportional to it.
The five-brane radius $r_{5,\, \mbox{cl}}$ 
can also be derived from the the classical five-brane action
\cite{Maldacena:2002rb}, 
\begin{eqnarray} 
r_{5,\, \mbox{cl}}^4 = \frac{1}{6} \frac{\mu N}{R}
\end{eqnarray} 
This calculation is valid for {\em large} values of $\lambda$.
measuring again in terms of $\frac{\mu}{R}$ we get 
\be
r_{5,\, \mbox{cl}}^2 \cdot \left(\frac{\mu}{R}\right)^{-2} = 
\frac{1}{9 \sqrt{2}} \sqrt{\lambda}
\ee  
This again depends only on $\lambda$, but has a different power dependence.
It is interesting to note that for the crossover region $\lambda \approx 1$,
$r_5$ and $r_{5,\, \mbox{cl}}$ are of the same order. It is natural to assume 
that there is a smooth transition between the small and large 't Hooft 
coupling results.

When comparing a single five-brane and a stack of $k$ of those, we would
like to deal with the same radius, $r_5$, and momentum per brane, $p^+$;
as we have seen, both must be large.
The equations (\ref{pplus},\ref{r5}) still apply, provided we replace 
$N$ by $n = N/k$ in them; $n$ is, so to speak, the amount of ``$N$ per brane''.
We therefore get that $\mu$ is fixed, and hence, as we have seen in this 
section, the Hagedorn temperature is fixed up to a very small numerical factor.
As we need $n$ fixed as we vary $k$, the total $N$ is proportional to $k$,
so we are not directly comparing different vacua of the matrix model with the 
same parameters. This result should be contrasted with the Little String Theory
case, arising from a stack of flat coincident NS five-branes in Type II string 
theory, where the Hagedorn temperature behaves 
\cite{Maldacena:1996ya,Maldacena:1997cg} as $\frac{1}{\sqrt{k}}$.
   
In addition, from subsection \ref{prefactor} we know that there is no 
polynomial prefactor for the exponential Hagedorn density of states.

For the membrane case, the constraints given by the validity domain of the 
approximations are much more lax, and our results are more robust. 
It is not entirely clear \cite{Dasgupta:2002hx} whether the corrections 
to the free, quadratic, results are governed by 
$\left(\frac{R}{\mu N}\right)^{3/2}$ or by 
$N^{1/2} \left(\frac{R}{\mu N}\right)^{3/2}$. In any case, that magnitude 
might be small while $R$ being either large or small in the 
$N \rightarrow \infty$ limit. 
The momentum of the membrane, $p^+$, and its (classical) radius
\be
r_2 = \frac{1}{6} \frac{\mu N}{R}
\ee 
should be large, although the ambiguity just mentioned does not allow us to    
decide whether they diverge in the large $N$ limit or whether they can be held
fixed.

\subsection{String Theory Perspective}

The DLCQ of M-theory, giving matrix models, can be thought of as the limit 
of quantizations of M-theory on nearly light-like circles, when an auxiliary
parameter $R_s$ with a dimension of length vanishes, $R_s \rightarrow 0$.   
By boosting, this parameter can be taken as the radius of a spatial 
compactification of another M-theory \cite{Sen:1997we,Seiberg:1997ad}.
This latter compactification gives rise to a type IIA string theory on a 
certain background.
Our aim in this subsection is to correlate the small $R_s$ limit
with the large $N$ limit so as to obtain 
a valid and illuminating string theory 
description of our system. 

The original M-theory background is a plane wave with parameter $\mu$,
and it is compactified on an (almost) light-like circle with radius $R$.
Performing an (almost) infinite boost, it can be regarded as
an M-theory background of a plane wave with a different parameter $\mu_s$,
compactified on a spatial circle with radius $R_s$.

More precisely, we start from a plane wave background
\begin{eqnarray}
ds^2&=&-2d{x'}^+d{x'}^- - \left( \left(\frac{\mu_s}{3}\right)^2(x^a)^2+
\left(\frac{\mu_s}{6}\right)^2(x^{a'})^2\right)({dx'}^+)^2+(dx^a)^2+(dx^{a'})^2
\nonumber \\
F_{+123}&=&\mu_s
\end{eqnarray}
with an identification
\be
\left(
 \begin{array}{c}
  t' \\
  {x'}^{10}
 \end{array}
\right)
\sim
\left(
 \begin{array}{c}
  t' \\
  {x'}^{10} + 2\pi R_s
 \end{array}
\right)
\ee
where
${x'}^{\pm} = \frac{1}{2} (t' + {x'}^{10})$.\footnote{
This convention is 
different from that of the other sections by a factor of $\sqrt{2}$,
which is not relevant for our rough estimates.}
This is related to our original background (\ref{pmetric})
by the boost
\be
\left(
 \begin{array}{c}
  t \\
  x^{10}
 \end{array}
\right)
=
\left(
 \begin{array}{c}
  t' \cosh \alpha  -   {x'}^{10}\sinh \alpha \\
  - t' \sinh \alpha  +   {x'}^{10}\cosh \alpha 
 \end{array}
\right)
\ee
or
\be
x^+ = e^{-\alpha} {x'}^+, 
x^- = e^{\alpha} {x'}^-
\ee
and the plane wave parameters are related as
\be
\label{mu}
\mu = e^\alpha \mu_s
\ee
The identification in the $(t, x^{10})$ coordinates is
\begin{eqnarray}
\left(
 \begin{array}{c}
  t \\
  x^{10}
 \end{array}
\right)
&\sim&
\left(
 \begin{array}{c}
  t - 2\pi R_s \sinh \alpha\\
  x^{10} + 2\pi R_s \cosh \alpha
 \end{array}
\right) \nonumber \\
&=&
\left(
 \begin{array}{c}
  t \\
  x^{10}
 \end{array}
\right)
+
2 \pi R
\left(
 \begin{array}{c}
  -1 \\
  1
 \end{array}
\right) + {\cal O} (e^{-\alpha})
\end{eqnarray}
where
\be
 \label{RRs}
R = e^\alpha R_s
\ee
Thus the light-like compactification
$x^- \sim x^- + 2\pi R $ 
with a plane wave parameter $\mu$
can be regarded as a limit of the 
spatial compactification ${x'}^{10} \sim {x'}^{10} + 2\pi R_s  $
in the infinitely boosted frame
$\alpha \rightarrow \infty$.
In this infinite boost limit,
$R_s$ is taken to be zero as $R_s \sim e^{-\alpha}$ so that
$R$ remains finite.

From (\ref{mu}) and (\ref{RRs}), and reinstituting the Planck energy scale 
in the original M-theory, we obtain
\be
\label{boost}
\frac{R   M_p^2}{\mu  } = 
\frac{R_s M_p^2}{\mu_s} 
\ee
in the $\alpha \rightarrow \infty$ limit.
Instead of changing $\mu$ to $\mu_s$, we prefer, following 
\cite{Susskind:1997cw}-\cite{Seiberg:1997ad},
to rescale the Planck length. 
Put otherwise, we map the original M-theory to a 
new M-theory, compactified on a spatial circle with radius $R_s$,
having a plane wave parameter $\mu$ (not $\mu_s$), and a Planck energy scale
$\tilde{M}_p$.
This new theory is denoted by $\tilde{\mbox{M}}$, 
and all of its parameters will be
written with a tilde. 
The advantage in doing so is that the light cone energy scale of the 
excitations in the original M-theory (which is $\mu$ in our case) 
is kept fixed under this map.
Explicitly, we demand 
\be
\label{boost2}
\frac{R_s \tilde{M}_p^2}{\mu  } \equiv
\frac{R           M_p^2}{\mu  } = 
\frac{R_s         M_p^2}{\mu_s}
\ee
and therefore obtain that
\be 
\label{tildeMp}
\frac{M_p^2}{\mu_s} = \frac{\tilde{M}_p^2}{\mu} 
\ee
or, from (\ref{mu}), that
$\tilde{M}_p^2 = e^{\alpha} M_p^2$. 
To keep $M_p$ finite and fixed, we need 
$\tilde{M}_p \rightarrow \infty$. 
We also rewrite (\ref{boost2}) as 
\be
R_s \tilde{M}_p^2 = R M_p^2
\ee

The $\tilde{\mbox{M}}$ theory is 
described by $N$ D0-branes in
weakly coupled type IIA string theory,
with \cite{Seiberg:1997ad} the string coupling
\footnote{These relations for flat space 
are valid only for a weak background.
This will turn out to be the case,
see below.
}
\be
\label{gs}
\tilde{g}_s = R_s^{3/4} (R M_p^2)^{3/4} \rightarrow 0
\ee
and string energy scale
\be
\label{Ms}
\tilde{M}_s = R_s^{-1/4} (R M_p^2)^{3/4} 
\ee
The 't Hooft coupling for the $\tilde{\mbox{M}}$ theory
is seen to be the same as the one in the M-theory,
\be
\label{lambda}
\lambda 
= \left(\frac{3 R M_p^2}{\mu}\right)^3 N
= \left(\frac{3 R_s M_p^2}{\mu_s}\right)^3 N
= \left(\frac{3 R_s \tilde{M}_p^2}{\mu}\right)^3 N
\ee
and in string units it is expressed as
\be
\label{stlmd}
\lambda =\tilde{g}_s N  \left(\frac{3 \tilde{M}_s}{\mu}\right)^3
\ee
In the standard prescription
of DLCQ in refs.\ \cite{Susskind:1997cw}-\cite{Seiberg:1997ad},
$R_s$ is simply taken to be zero.
We wish to consider a version
of DLCQ which is appropriate for our 't Hooft limit.
We scale $R_s$ in an $N$ dependent way
in the large $N$ limit,
so that
the 't Hooft coupling $\tilde{g}_s N$
of open string theory on D0-branes is kept finite.
From (\ref{stlmd}) this means 
$
\tilde{M}_s/{\mu}
$
is kept finite, which is different from the
case studied in refs.\ \cite{Susskind:1997cw}-\cite{Seiberg:1997ad}.
Later we will see, however, that although finite
the string energy scale should be kept large
compared with $\mu$.
Then from (\ref{Ms})
\be
\left(\frac{\tilde{M}_s}{\mu}\right)
\sim
\lambda^{1/4} (\mu N R_s)^{-1/4}
\ee
so we take the following large $N$ scaling of $R_s$
to keep 
$\tilde{M}_s/{\mu}$ finite,
\be
R_s \sim \frac{\lambda}{\mu N} \left(\frac{\mu}{\tilde{M}_s}\right)^4
\ee
This is equivalent to taking the 
boost parameter as
\be
e^{-\alpha}=\frac{R_s}{R}
\sim
\frac{\lambda^{2/3}M_p^2}{\mu^2}
\left(\frac{\mu}{\tilde{M}_s}\right)^4
N^{2/3}
\ee
From (\ref{lambda}) we also need to take
\be
\tilde{M}_p \sim \mu \lambda^{-1/3} N^{1/3} 
\left(\frac{\mu}{\tilde{M}_s}\right)^{-2}\rightarrow \infty
\ee
The radius $r_5$ of the spherical five-brane 
(\ref{r5}) is also rescaled, as a transverse direction, in the new theory,
\be
\tilde{r}_5 \tilde{M}_p  = r_5 M_p
\ee
Then
\be
\tilde{r}_5^2 = \frac{M_p^2}{\tilde{M}_p^2} r_5^2 
\sim \frac{R_s N}{\mu}
\sim \frac{\lambda }{\mu^2} \left(\frac{\mu}{\tilde{M}_s}\right)^4
\ee
Measuring the rescaled radius in the string
units, we obtain
\be
 \label{r5st}
\tilde{r}_5^2 \tilde{M}_s^2 \sim
\lambda  \left(\frac{\mu}{\tilde{M}_s}\right)^2
\ee
Thus the radius of the transverse spherical M five-brane,
or equivalently 
spherical NS five-brane in IIA string theory,
is kept finite and small in string units.

We need to check the validity
of the flat space formulas (\ref{gs}) and (\ref{Ms})
we have used.
The plane wave metric in $(t, x^{10})$ coordinate is
\be
\label{pspace}
ds^2 = 
-\frac{1}{2}(1+\frac{F^2}{4}) dt^2 + \frac{1}{2}(1-\frac{F^2}{4}) (dx^{10})^2 
-\frac{F^2}{4}  dt dx^{10} 
+ (dx^a)^2+(dx^{a'})^2, 
\ee
where
\be
F^2 = \left(\frac{\mu}{3}\right)^2(x^a)^2+
\left(\frac{\mu}{6}\right)^2(x^{a'})^2
\ee
Thus the flat space relations (\ref{gs}), (\ref{Ms}) are valid when the 
transverse length scales we are considering are 
much smaller than $1/\mu$, thus we need
\be
\mu \tilde{r}_2, \, \mu \tilde{r}_5 \ll 1
\ee
where
\be
\tilde{r}_2 \sim \frac{\mu}{R_s \tilde{M}_p^3} 
\ee
is the size of the fuzzy membrane solution.
This is the case when the 't Hooft coupling is finite
and 
\be
\mu \tilde{r}_2
\sim
\left(\frac{\mu}{\tilde{M}_s}\right)^2 \ll 1
\ee
i.e.
$\mu/\tilde{M}_s$ is kept finite and small.
This condition justifies the 
use of the D0-brane action without stringy correction.

We conclude that our 't Hooft limit is
the limit where the string scale and
the radius of the spherical NS five-brane are fixed.
String theory in an NS five-brane
background with a fixed string scale gives
the little string theory (LST).
The validity of our calculations
is limited to small $\lambda$ 
where the NS five-brane is highly curved
in string units, and this prevents
us from direct comparison with the results
of LST on flat space \cite{Kutasov:2001uf}.
However, the existence of the phase transition appears
robust and is expected to occur over a range of $\lambda$ values.

To recapitulate, the original M-theory compactification has the following 
parameters: the Planck energy scale $M_p$, $N$, the null circle radius $R$,
and the plane wave parameter $\mu$. We look at the compactification as the 
limit $R_s \rightarrow 0$ of nearly light-like compactifications, and map,
via a boost, to a string theory with the following parameters: the string
energy scale $\tilde{M}_s$, $N$, the string coupling $\tilde{g}_s$ and $\mu$.
Taking the $R_s \rightarrow 0$ limit in a certain way, we are able to satisfy
the assumptions underlying our approximations: in the M-theory side, those are
the large $N$ limit, $N \rightarrow \infty$, and the weakly coupled 
matrix model condition $\lambda \ll 1$; we take the 't Hooft limit of fixed 
$\lambda$. 
In the string theory side, there are the decoupling limit condition 
$\tilde{g}_s \rightarrow 0$ and the weak background condition 
$\mu \ll \tilde{M}_s$ 
near the origin of the space, where the branes reside. 
We then conclude from (\ref{r5st}) that the NS five-brane size is 
finite and small in string units, 
\be
\tilde{r}_5 \tilde{M}_s \ll 1
\ee

\subsection{Outlook}

If we do not fix the light-cone momentum,
the finite temperature partition function of the matrix model should
use the rest frame energy
\begin{equation}
p^0=\frac{1}{\sqrt{2}}\left( p^++p^-\right)
\end{equation}
in the Boltzmann weight for the partition function,
\begin{equation}
Z=\tr e^{-\frac {\beta} {\sqrt{2}} \left( p^++p^-\right) }
\end{equation}
The trace over $p^+$ is a sum over $N$ and the trace over the states of
$p^-$ is the partition function of the matrix model,
\begin{equation}
Z=\sum_{N=0}^\infty e^{-\frac{\beta N}{\sqrt{2}R}}e^{-\beta F[\beta, N]}
\end{equation}
Here, $F[\beta,N]$ is the free energy of the matrix model.
The convergence of this sum for large $N$ is determined by the free energy
in the large $N$ limit.

In the membrane states, the free energy is of order one in the large N limit
and the partition sum converges.  In this same limit, the M five-brane states
are strongly coupled unless we take the 't Hooft limit.  In that case, there
is a phase transition and the behavior of the free energy changes from being
of order one to order $N^2$.  Further, when it is of order $N^2$, it is
negative and the partition sum diverges.

This divergence could be interpreted as an instability of the finite temperature system.  The logical speculation is the collapse of the system to a black hole.
This is certainly what we would expect in finite temperature field theory on a
Minkowski background.  There, we would expect that the Jeans instability renders
any theory of quantum gravity unstable at finite temperature.  Indeed, in the
BFSS matrix model, we do expect the free energy to grow faster than linearly in
$N$.

\appendix
\section{Fixing the Static Diagonal Gauge}

The thermodynamic partition function is defined by the Euclidean
path integral
\begin{equation}
Z=\int [dA]...\exp\left( -\int_0^{1/T}d\tau
\frac{1}{2}{\rm Tr}\left( (DX)^2+\ldots \right)\right)
\label{partf}
\end{equation}
where $T$ is the temperature (we use units where Boltzmann's constant $k_B=1$)
the variables have periodic boundary conditions,
$$
A(\tau)=A(\tau+1/T)
~~,~~
X(\tau)=X(\tau+1/T)
~~,~~\ldots
$$

This model has the gauge invariance,
$$
X\to U(t)X(t)U^{\dagger}(t)
~~,~~
A(t)\to U(t)\left( A(t)-i\frac{d}{dt}\right)U^{\dagger}(t)
$$
where $U(\tau)=zU(\tau+1/T)$ is a periodic function of time and $z$ is constant element of the center of the group.   It is present here since the gauge transform is entirely in the adjoint.  In a sense, the true gauge group is the factor
group $U(N)/Z_N$.  There is a quantity, the Polyakov loop operator
$$
P={\rm Tr}\left( {\cal P}e^{i\int_0^{1/T}d\tau A(\tau)}\right)
$$
which transforms non-trivially,
$$
P \to zP
$$
This is interpreted as a global symmetry of the theory.
Its realization is related to the gauge interaction in the theory and
can be used to analyze confinement.  There are two phases.  
When the symmetry is good, the Polyakov loop operator averages to zero:
$
\left< P \right> = 0
$.  When the symmetry is spontaneously broken, the Polyakov loop operator
can have non-vanishing expectation value, $\left< P \right> \neq 0$.  
Of course in this very low dimensional system, a discrete symmetry can only 
be broken in the large $N$ limit.

The expectation value of $P$ is interpreted as the energy of this system
with one additional external fundamental representation source.  Indeed, the
path integral quantization of the system with such a source would replace the
partition function (\ref{partf}) by $\left< P \right>$ and with $n$ such 
sources the partition function would be replaces by $\left< P^n \right>$.
Thus, we interpret
$$
F[T]=-T\ln\left(\left< P \right>\right)
$$
as the free energy that it would take to  introduce a fundamental representation
source into the system.  When $\left< P \right>=0$ this free energy is 
infinite and we say that the system is confining, whereas when 
$\left< P \right>\neq 0$ this symmetry is finite and we
say that the system is deconfined.

Let us analyze the partition function (\ref{partf}) in more detail.
We can fix a static gauge by enforcing the gauge condition
$$
\frac{d}{d\tau}A(\tau)=0
$$
Then, taking into account the Faddeev-Popov ghost determinant, the
partition function becomes
$$
Z=\int [dA(\tau)]\ldots \delta\left(dA/d\tau\right)
\det \left[-\frac{d}{d\tau}D_\tau\right]
\exp\left( -\int_0^{1/T}d\tau
\frac{1}{2}{\rm Tr}\left( (DX)^2+\ldots \right)\right)
$$
The measure in the path integral can be defined using the Fourier transform,
$$
A(\tau)=\sqrt{ T} \sum_{n=-\infty}^\infty A_n e^{2\pi i n T \tau}
$$
The integration measure can be defined in this basis,
$$
[dA(\tau)]\equiv \prod_n dA_n
$$

The Faddeev-Popov determinant is a determinant over
the non-zero modes,
$$
\delta\left(d A(\tau)/d\tau\right)
\det \left[-\frac{d}{d\tau}D_\tau\right]
=
\prod_{a,b}\prod_{n\neq 0}\delta\left(2\pi i nTA_n\right)
\cdot\left[(2\pi i nT) (2\pi i nT +iA^{\rm adj}_{ab}) \right]
$$
$$
=\prod_{a,b}\prod_{n\neq 0}\delta\left( A_n\right)
\cdot\left[ (2\pi i nT +iA^{\rm adj}_{ab})\right]
$$
where $A^{\rm adj}_{ab}$ is the adjoint.
Integrating the modes of $A_n$ with $n\neq 0$ and
decomposing the measure into an integral over the eigenvalues of $A$ and the
unitary matrices necessary to diagonalize yields the measure
$$
[dA]=\prod_{a=1}^N dA_a \prod_{b\neq a}(A_a-A_b)^2[d\Omega]
$$
where $\Omega$ is the unitary matrix which diagonalizes $A$.
Also, when $A$ is diagonal $A^{\rm adj}_{ab}=A_a-A_b$.
Then, the total measure becomes
$$
\prod_{a=1}^N d A_a \prod_{b\neq a}\prod_{n=-\infty}^\infty
(2\pi i nT +i(A_a-A_b))\left(\prod_{n\neq 0}2\pi i nT\right)^N
$$
$$
=\prod_{a=1}^N \frac{d A_a}{T} \prod_{b\neq a}\prod_{n=-\infty}^\infty
(2\pi i nT +i(A_a-A_b))
$$
where we have used the zeta-function regularization
to evaluate the infinite product
$$
\prod_{n\neq 0}2\pi i nT= \prod_{n=1}^\infty(2\pi n T)^2
=(2\pi  T)^{2\zeta(0)}e^{-2\zeta'(0)} =\frac{1}{T}
$$
Riemann's zeta function is defined by
$$
\zeta(s)=\sum_{n=1}^\infty \frac{1}{n^s}
$$
and has the values
$$
\zeta(0)=-\frac{1}{2}
~~,~~
\zeta'(0)=-\frac{1}{2}\ln(2\pi)
$$

To take the remaining product, we use the
formula
$$ \prod_{n=-\infty}^\infty
(2\pi i nT +i\phi)=\exp\left(\sum_n \ln(2\pi i n T+i\phi)\right)
$$
$$
=\exp\left( -\int d\phi \int d\phi \sum\frac{1}{(2\pi i n T
+i\phi)^2}\right)
$$
$$
=\exp\left( -\int d\phi \int d\phi \frac{1}{2\pi i}\oint_C dz
\frac{1}{2T}\coth (z/2T) \frac{1}{(z+i\phi)^2}\right)
$$
where we have taken two derivatives in order to get a convergent series and
the contour $C$ encircles all of the singularities of the $\coth z/2T $ function
located at $z=2\pi i nT$.
Doing the contour integral by collapsing the contour onto the singularity at
$z=-i\phi$ gives
$$ \prod_{n=-\infty}^\infty
(2\pi  inT +i\phi) =
\exp\left( -\int d\phi \int d\phi \frac{d}{d(-i\phi)}
\frac{1}{2T}\coth (-i\phi/2T) \right)
$$
$$
=
\exp\left( \int d\phi\left(
\frac{1}{2T}\cot (\phi/2T)+\beta\right) \right)
=\exp\left( \ln \sin(\phi/2T)+\alpha+\beta\phi\right)=
e^{\alpha+\beta\phi}\sin(\phi/2T)
$$
where $\alpha$ and $\beta$ are arbitrary constants parameterizing the subtractions
that are needed to renormalize the product.
If we {\it arbitrarily} set both of them to zero, we get the gauge-fixed
measure for integrating over gauge fields as
$$
\prod_{a=1}^N \frac{dA_a}{T} \prod_{b\neq a}\left|
\sin\left( \frac{A_a-A_b}{2}\right)\right|
$$
The absolute value sign is there because the Faddeev-Popov determinants should always have absolute value signs.

\setlength{\baselineskip}{0.666666667\baselineskip}


\begin{thebibliography}{99}


\bibitem{Banks:1996vh}
T.~Banks, W.~Fischler, S.~H.~Shenker and L.~Susskind,
``M theory as a matrix model: A conjecture,''
Phys.\ Rev.\ D {\bf 55}, 5112 (1997)
[arXiv:hep-th/9610043].

\bibitem{Taylor:2001vb}
W.~Taylor,
``M(atrix) theory: Matrix quantum mechanics as a fundamental theory,''
Rev.\ Mod.\ Phys.\  {\bf 73}, 419 (2001)
[arXiv:hep-th/0101126].

\bibitem{deWit:1988ig}
B.~de Wit, J.~Hoppe and H.~Nicolai,
``On The Quantum Mechanics Of Supermembranes,''
Nucl.\ Phys.\ B {\bf 305}, 545 (1988).



\bibitem{Berenstein:2002jq}
D.~Berenstein, J.~M.~Maldacena and H.~Nastase,
``Strings in flat space and pp waves from N = 4 super Yang Mills,''
JHEP {\bf 0204}, 013 (2002)
[arXiv:hep-th/0202021].


\bibitem{Dasgupta:2002hx}
K.~Dasgupta, M.~M.~Sheikh-Jabbari and M.~Van Raamsdonk,
``Matrix perturbation theory for M-theory on a PP-wave,''
JHEP {\bf 0205}, 056 (2002)
[arXiv:hep-th/0205185].


\bibitem{Dasgupta:2002ru}
K.~Dasgupta, M.~M.~Sheikh-Jabbari and M.~Van Raamsdonk,
``Protected multiplets of M-theory on a plane wave,''
JHEP {\bf 0209}, 021 (2002)
[arXiv:hep-th/0207050].

\bibitem{Kim:2002if}
N.~w.~Kim and J.~Plefka,
``On the spectrum of pp-wave matrix theory,''
Nucl.\ Phys.\ B {\bf 643}, 31 (2002)
[arXiv:hep-th/0207034].

\bibitem{Myers:1999ps}
R.~C.~Myers,
``Dielectric-branes,''
JHEP {\bf 9912}, 022 (1999)
[arXiv:hep-th/9910053].

\bibitem{Polchinski:2000uf}
J.~Polchinski and M.~J.~Strassler,
``The string dual of a confining four-dimensional gauge theory,''
arXiv:hep-th/0003136.

\bibitem{Maldacena:2002rb}
J.~Maldacena, M.~M.~Sheikh-Jabbari and M.~Van Raamsdonk,
``Transverse fivebranes in matrix theory,''
JHEP {\bf 0301}, 038 (2003)
[arXiv:hep-th/0211139].



\bibitem{Sundborg:1999ue}
B.~Sundborg,
``The Hagedorn transition, deconfinement and N = 4 SYM theory,''
Nucl.\ Phys.\ B {\bf 573}, 349 (2000)
[arXiv:hep-th/9908001].

\bibitem{Polyakov:2001af}
A.~M.~Polyakov,
``Gauge fields and space-time,''
Int.\ J.\ Mod.\ Phys.\ A {\bf 17S1}, 119 (2002)
[arXiv:hep-th/0110196].

\bibitem{AMvR:seminars}
S.~Minwalla,
talk at ``Recent Developments in String Theory'', Banff, March 2003;\\
O.~Aharony, 
seminar at the University of British Columbia, Vancouver, September 2003;\\
M.~Van~Raamsdonk, 
private communication.

\bibitem{Edwards}
G.~E.~Andrews,
``The Theory of Partitions'', Encyclopedia of Mathematics and its Applications,
volume 2, Addison--Wesley Publishing Company, 1976.

\bibitem{Maldacena:1997re}
J.~M.~Maldacena,
``The large N limit of superconformal field theories and supergravity,''
Adv.\ Theor.\ Math.\ Phys.\  {\bf 2}, 231 (1998)
[Int.\ J.\ Theor.\ Phys.\  {\bf 38}, 1113 (1999)]
[arXiv:hep-th/9711200].

\bibitem{Maldacena:1996ya}
J.~M.~Maldacena,
``Statistical Entropy of Near Extremal Five-branes,''
Nucl.\ Phys.\ B {\bf 477}, 168 (1996)
[arXiv:hep-th/9605016].

\bibitem{Maldacena:1997cg}
J.~M.~Maldacena and A.~Strominger,
``Semiclassical decay of near-extremal fivebranes,''
JHEP {\bf 9712}, 008 (1997)
[arXiv:hep-th/9710014].

\bibitem{Kutasov:2001uf}
See, e.g. D.~Kutasov,
``Introduction to little string theory,''
(prepared for the ICTP Spring School on Superstrings and Related Matters, 
Trieste, Italy, 2-10 April 2001),
and references therein.



\bibitem{Susskind:1997cw}
L.~Susskind,
``Another conjecture about M(atrix) theory,''
arXiv:hep-th/9704080.

\bibitem{Sen:1997we}
A.~Sen,
``D0 branes on $T^n$ and matrix theory,''
Adv.\ Theor.\ Math.\ Phys.\  {\bf 2}, 51 (1998)
[arXiv:hep-th/9709220].

\bibitem{Seiberg:1997ad}
N.~Seiberg,
``Why is the matrix model correct?,''
Phys.\ Rev.\ Lett.\  {\bf 79}, 3577 (1997)
[arXiv:hep-th/9710009].


\bibitem{Ambjorn:1998zt}
J.~Ambjorn, Y.~M.~Makeenko and G.~W.~Semenoff,
``Thermodynamics of D0-branes in matrix theory,''
Phys.\ Lett.\ B {\bf 445}, 307 (1999)
[arXiv:hep-th/9810170].

\bibitem{Boulatov:1991fp}
D.~Boulatov and V.~Kazakov,
``Vortex anti-vortex sector of one-dimensional string theory via the upside down matrix oscillator,''
Nucl.\ Phys.\ Proc.\ Suppl.\  {\bf 25A}, 38 (1992).

\bibitem{Douglas:2003up}
M.~R.~Douglas, I.~R.~Klebanov, D.~Kutasov, J.~Maldacena, E.~Martinec and N.~Seiberg,
``A new hat for the c = 1 matrix model,''
arXiv:hep-th/0307195.


\bibitem{Gross:he}
D.~J.~Gross and E.~Witten,
``Possible Third Order Phase Transition In The Large N Lattice Gauge Theory,''
Phys.\ Rev.\ D {\bf 21}, 446 (1980).

\bibitem{Hyun:2002wp}
S.~j.~Hyun and H.~j.~Shin,
``Solvable N = (4,4) type IIa string theory in plane-wave background and D-branes,''
Nucl.\ Phys.\ B {\bf 654}, 114 (2003)
[arXiv:hep-th/0210158].

\bibitem{Hyun:2002wu}
S.~j.~Hyun and H.~j.~Shin,
``N = (4,4) type IIA string theory on pp-wave background,''
JHEP {\bf 0210}, 070 (2002)
[arXiv:hep-th/0208074].

\bibitem{Hyun:2002cm}
S.~j.~Hyun and H.~j.~Shin,
``Branes from matrix theory in pp-wave background,''
Phys.\ Lett.\ B {\bf 543}, 115 (2002)
[arXiv:hep-th/0206090].



\bibitem{Das:2003yq}
S.~R.~Das, J.~Michelson and A.~D.~Shapere,
``Fuzzy spheres in pp-wave matrix string theory,''
arXiv:hep-th/0306270.


\bibitem{Grignani:2003cs}
G.~Grignani, M.~Orselli, G.~W.~Semenoff and D.~Trancanelli,
``The superstring Hagedorn temperature in a pp-wave background,''
JHEP {\bf 0306}, 006 (2003)
[arXiv:hep-th/0301186].

\bibitem{Grignani:2000zm}
G.~Grignani, P.~Orland, L.~D.~Paniak and G.~W.~Semenoff,
``Matrix theory interpretation of DLCQ string worldsheets,''
Phys.\ Rev.\ Lett.\  {\bf 85}, 3343 (2000)
[arXiv:hep-th/0004194].

\bibitem{Grignani:1999sp}
G.~Grignani and G.~W.~Semenoff,
``Thermodynamic partition function of matrix superstrings,''
Nucl.\ Phys.\ B {\bf 561}, 243 (1999)
[arXiv:hep-th/9903246].

\bibitem{Grignani:1995hx}
G.~Grignani, G.~W.~Semenoff, P.~Sodano and O.~Tirkkonen,
``Charge Screening and Confinement in the Hot 3D-QED,''
Nucl.\ Phys.\ B {\bf 473}, 143 (1996)
[arXiv:hep-th/9512048].

\bibitem{Semenoff:1996ew}
G.~W.~Semenoff and K.~Zarembo,
``Adjoint non-Abelian Coulomb gas at large N,''
Nucl.\ Phys.\ B {\bf 480}, 317 (1996)
[arXiv:hep-th/9606117].

\bibitem{Semenoff:1996xg}
G.~W.~Semenoff, O.~Tirkkonen and K.~Zarembo,
``Exact solution of the one-dimensional non-Abelian Coulomb gas at  large N,''
Phys.\ Rev.\ Lett.\  {\bf 77}, 2174 (1996)
[arXiv:hep-th/9605172].

\bibitem{AMvR:paper}
O.~Aharony, J.~Marsano, S.~Minwalla, K.~Papadodimas and M.~Van~Raamsdonk,
``The Hagedorn/Deconfinement Phase Transition in Weakly Coupled Large $N$ 
Gauge Theories,''
arXiv:hep-th/0310285.


\end{thebibliography}
\end{document}